\documentclass[12pt]{iopart}
\pdfoutput=1 
\usepackage{graphicx}
\usepackage{latexsym}
\usepackage[T1]{fontenc}
\usepackage{color}
\usepackage{amssymb}
\usepackage{endnotes}
\usepackage{lmodern}
\usepackage{float}
\usepackage{color}
\newcommand{\red}{\color{black}}
\usepackage[rightcaption]{sidecap}

\begin{document}
	
\JPCM

\topical{Nematicity, magnetism and superconductivity in FeSe}

\author{Anna E B\"ohmer\footnote{Now at Institut f\"ur Festk\"orperphysik, Karlsruhe Institute of Technology, D-76021 Karlsruhe, Germany}}
\address{Ames Laboratory, US DOE, Ames, IA 50011, USA}

\author{Andreas Kreisel}
\address{Institut f\" ur Theoretische Physik, Universit\" at Leipzig, D-04103 Leipzig, Germany}

\begin{abstract}
Iron-based superconductors are well known for their complex interplay between structure, magnetism and superconductivity. FeSe offers a particularly fascinating example. This material has been intensely discussed because of its extended nematic phase, whose relationship with magnetism is not obvious. Superconductivity in FeSe is highly tunable, with the superconducting transition temperature, $T_\mathrm{c}$, ranging from 8 K in bulk single crystals at ambient pressure to almost 40 K under pressure or in intercalated systems, and to even higher temperatures in thin films. In this topical review, we present an overview of nematicity, magnetism and superconductivity, and discuss the interplay of these phases in FeSe. We focus on bulk FeSe and the effects of physical pressure and chemical substitutions as tuning parameters. The experimental results are discussed in the context of the well-studied iron-pnictide superconductors and interpretations from theoretical approaches are presented. 
\end{abstract}

\setcounter{tocdepth}{3}
\tableofcontents

\section{Introduction}

Unconventional superconductivity in strongly correlated materials often {\red emerges close to} phases characterized by magnetic, structural, orbital, or other types of order. A change of tuning parameters, like external pressure or the chemical composition, can impact these phases and induce transitions between them, resulting in complex phase diagrams. Iron-based superconductors are an ideal platform to study such interacting phases. These intriguing materials feature a complex interplay of high-temperature superconductivity, antiferromagnetic orders, structural distortions and orbital order. The large variety of available materials and suitable tuning parameters offers numerous possibilities for systematic and comparative studies. FeSe represents an unusual and particularly fascinating case of the interplay between structural, magnetic and electronic degrees of freedom.

\subsection{Phase interplay in iron pnictides}

\begin{figure}
	\includegraphics[width=\textwidth]{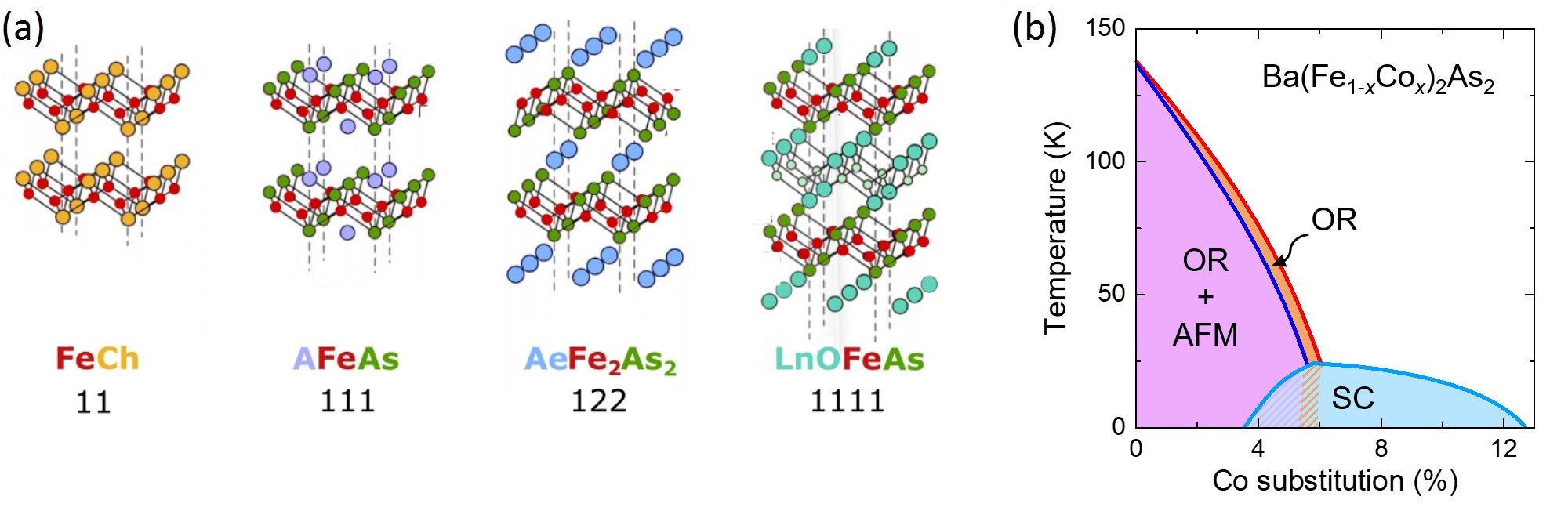}
	\caption{Crystal structures and typical phase diagram of iron-based superconductors. (a) Atomic structure of the four most common iron-based material families. In the 11-system, the chalcogenide atom Ch can be S, Se or Te. Representatives of the 111-type material family are LiFeAs and NaFeAs. In the 122-systems $Ae$Fe$_2$As$_2$, $Ae$ stands for Ca, Sr, Ba or Eu. Numerous substitution series (e.g., Co, Ni, Ru, Rh, Pd, Pt or Ir for Fe; Na, K, Rb for $A$; or P for As) display superconductivity. In the 1111-system, Ln stands for a lanthanide ion, $Ln$=La, Pr, Nd, Sm, Gd or Tb and a similarly wide variety of substitutions induces superconductivity. Partial substitution of O with F or H are particularly common. Reproduced with permission from \cite{Boehmer2017_phiuz}. \copyright\ 2017 Wiley-VCH Verlag GmbH \& Co. KGaA, Weinheim. (b) A schematic phase diagram of Co-substituted BaFe$_2$As$_2$, featuring an, orthorhombic (OR) and paramagnetic (nematic) region (orange), an orthorhombic and antiferromagnetic (AFM) region (rose) and a superconducting (SC) dome (blue).
	\label{fig:1}}
\end{figure}

Two distinctive features of iron-based materials are their complex phase interplay and a surprisingly large chemical variety. After the discovery of superconductivity in F-doped LaFeAsO\cite{Kamihara2008}, many related compounds were found. The materials are classified according to their stoichiometry into 1111-type, 122-type, 111-type, etc \cite{Paglione2010,Johnston2010,Stewart2011,Hosono2015a}, see Fig. \ref{fig:1} (a). 
The 122-type material family, with BaFe$_2$As$_2$ as its most prominent member, is arguably the most intensively studied. This is likely due to the relative ease of preparation of sizable single crystals and to an impressive variety of chemical substitutions that are capable of inducing superconductivity.  
These 122-type materials have shaped a picture of 'typical' interplay between orthorhombic distortion, nematicity, magnetism and superconductivity in iron-based materials (Fig. \ref{fig:1}(b)), into which other compounds, like NaFeAs\cite{Parker2010} or even the structurally very complex 10-3-8 system\cite{Sapkota2014} also fit well. The phase interplay of FeSe, however, does not seem to follow this pattern in several respects. 

As also observed in many other unconventional superconductors, the superconducting instability in iron-based materials typically emerges when antiferromagnetic order is sufficiently reduced by modifying a tuning parameter. The superconducting transition temperature $T_\mathrm{c}$ has then a dome-like dependence on this tuning parameter\cite{Uemura2009}. A particularity of the iron-based materials is that their antiferromagnetism is (with few exceptions)  ``stripe type'', i.e., it distinguishes two perpendicular in-plane directions. In consequence, magnetic order is intimately coupled to a an orthorhombic lattice distortion, whereas the paramagnetic phase is typically tetragonal. Simultaneous magnetic and structural phase transitions occur, for example, in (Ba$_{1-x}$K$_x$)Fe$_2$As$_2$\cite{Avci2012} and Ba(Fe$_{1-x}$Ru$_x$)$_2$As$_2$\cite{Thaler2010}. However, in a number of systems, the structural transition at $T_\mathrm{s}$ precedes the magnetic transition at $T_\mathrm{N}$ by several degrees upon cooling. Prominent examples are Co- or Ni-substituted BaFe$_2$As$_2$\cite{Ni2009,Canfield2010}, LaFeAsO\cite{Cruz2008} and NaFeAs\cite{Parker2010}, which thus host a paramagnetic orthorhombic phase. This observation sparked the idea that the structural transition is related to a distinct, ``nematic'' electronic degree of freedom. 

Nematicity has been a central theme in the study of iron-based materials\cite{Fernandes2014}. The term nematic is borrowed from the field of liquid crystals, where it refers to a phase with broken rotational but preserved translational symmetry\cite{Fernandes2012}. In the context of iron-based superconductors, ``nematicity'' and ``nematic order'' is often used synonymously to in-plane anisotropy, which means reduced rotational symmetry with respect to the tetragonal high-temperature phase. This electronic $a$-$b$ anisotropy can be observed in various experiments, including electronic transport \cite{Chu2010}, optical reflectivity \cite{Dusza2011}, angular-resolved photoemission spectroscopy (ARPES) \cite{Yi2011}, scanning tunneling microscopy (STM) \cite{Chuang2010}, nuclear magnetic resonance (NMR) \cite{Fu2012} and inelastic neutron scattering \cite{Lu2014}. 
In general, a macroscopic crystal forms fine scale structural twins below $T_\mathrm{s}$\cite{Tanatar2009} such that the macroscopic response of the crystal is isotropic. In order to study the in-plane anisotropy associated with the nematic phase by macroscopic probes, samples need to be detwinned. Detwinning is often accomplished by the application of uniaxial stress \cite{Fisher2011}. Microscopic or local probes like NMR, ARPES, diffraction, or STM can resolve the domain structure or detect signals from the two types of domains separately without the need for detwinning.
 
By symmetry, all properties---be they structural, orbital or magnetic---acquire in-plane anisotropy in the orthorhombic phase, which complicates the determination of the microscopic origin of nematicity in iron-based materials\cite{Fernandes2014}. 
This question may therefore only be solved by a combination of experiment and microscopic theories. Despite the separation of $T_\mathrm{s}$ and $T_\mathrm{N}$, the two transitions follow each other rather closely as function of tuning parameters in most systems, suggesting their intimate coupling. Notably, the intensity of magnetic fluctuations in NMR\cite{Ma2011,Ning2014}, as well as the spin-spin correlation length measured by neutron scattering\cite{Zhang2015_PRL} increase below $T_\mathrm{s}$. Reciprocally, the magnitude of structural distortion increases below $T_\mathrm{N}$\cite{Kim2011}. This close coupling between structure and magnetism supports the idea that nematicity may be consequence of magnetic interactions even when nematic order precedes magnetic order. In this so-called spin-nematic scenario \cite{Fernandes2012}, it is pointed out that stripe-type correlations between magnetic moments distinguish a direction in which the moments are primarily parallel from a direction in which they are antiparallel. These correlations, which break the tetragonal symmetry, can be long-range even if the magnetic moments themselves are not ordered. Thus, in this scenario, the tetragonal-to-orthorhombic transition can be induced by magnetic fluctuations at a temperature $T_\mathrm{s}>T_\mathrm{N}$.  

Superconductivity is observed when the structural and magnetic transitions are sufficiently suppressed by a chemical substitution or pressure. It often coexists and competes with these phases in a part of the phase diagram (see Fig. \ref{fig:1}(b)), as shown by a significant decrease in magnetic and structural (nematic) order parameters below $T_\mathrm{c}$\cite{Fernandes2010II,Nandi2010}. Superconductivity is likely mediated by magnetic fluctuations\cite{Mazin2009}.

Thus, in the general picture of phase interplay in iron-based materials, a stripe-type antiferromagnetic phase is intimately coupled to a nematic phase characterized by a sizable electronic $a$-$b$ anisotropy and a structural orthorhombic distortion. Superconductivity competes with both nematic and magnetic order.

\subsection{FeSe and related systems}  

FeSe is highly interesting in the context of phase interplay in iron-based materials because it hosts apparently very unusual interrelations between nematicity, magnetism and superconductivity. Bulk FeSe undergoes a tetragonal-to-orthorhombic transition at $T_\mathrm{s}\approx90$ K \cite{Hsu2008,Margadonna2008,McQueen2009} similarly to the ``nematic'' transition of many other iron-based parent materials. However, no magnetic order is formed at ambient pressure \cite{McQueen2009,Bendele2010} and FeSe is superconducting below $\sim8$ K \cite{Hsu2008}. Under the application of pressure, $T_\mathrm{s}$ decreases \cite{Miyoshi2014}, a magnetically ordered phase emerges at $\sim 1$ GPa \cite{Bendele2010,Bendele2012} and $T_\mathrm{c}$ increases dramatically to a maximum of $\approx37$ K\cite{Mizuguchi2008,Medvedev2009,Margadonna2009,Garbarino2009,Masaki2009,Okabe2010} at $\sim 6$ GPa.
 
Chemical substitutions can be seen as tuning parameters for bulk FeSe having partially similar effects as external pressure. Te or S substitution for Se are most commonly studied\cite{Mizuguchi2009}. The endmembers of these solid solutions, FeTe and FeS, are interesting materials in themselves\cite{Bao2009, Lai2015}. The structurally analogous FeTe$_{0.8}$S$_{0.2}$ material can even be tuned to superconductivity by simmering in alcoholic beverages\cite{Deguchi2012}. Intercalation of FeSe is possible with various elements and molecules\cite{Burrard-Lucas2013,Ying2011II} leading to an enhancement of $T_\mathrm{c}$ up to $\sim45$ K. In addition, the superconducting phase in the K-Fe-Se system appears to be closely related and may be viewed as K-intercalated FeSe\cite{Krzton-Maziopa2016}. Finally, thin films of FeSe are an intriguing topic. The single-layer FeSe on SrTiO$_3$\cite{Sadovskii2016,Wang2017,Huang2017} has surprised with the highest $T_\mathrm{c}$ of all iron-based materials, possibly reaching over 100 K\cite{Ge2015}. Multilayer thin films, on the other hand, seem to have 'conventional' properties and resemble bulk FeSe\cite{Song2011II}. 

Superconductivity in the layered PbO-type structure of FeSe (space group P4/nmm) was reported in 2008\cite{Hsu2008} just two months after the report of superconductivity in the 122-type systems. However, the complex phase interplay in FeSe was only gradually revealed during the subsequent years. This is primarily due to difficulties in obtaining phase-pure and high-quality single crystals. Single-crystal growth of FeSe is made difficult by the complex binary Fe-Se composition-temperature phase diagram\cite{Okamoto1991}. In particular, the superconducting tetragonal PbO-type and near-stoichiometric phase of FeSe has only a very narrow range of stability and undergoes a transformation to a hexagonal NiAs-type phase on warming above 457$^\circ$C. In consequence, any preparation procedure above this temperature results in samples that have not formed in the tetragonal phase. They often have hexagonal morphology and consist of a mix of tetragonal and hexagonal phases\cite{Braithwaite2009}, which inevitably leads to internal strains.

Early studies of FeSe used polycrystalline material prepared by solid state synthesis\cite{Hsu2008,Mizuguchi2008,Pomjakushina2009,McQueen2009II,Margadonna2009,Kumar2010}, containing the hexagonal NiAs-type impurity phase. Furthermore, the tetragonal FeSe phase has a finite width of formation and the superconducting transition temperature is very sensitive to the exact stoichiometry\cite{McQueen2009II}. Many early studies were performed on such polycrystalline, multi-phase material. There are also several early studies of the growth of tetragonal FeSe using Cl-salt-based flux techniques\cite{Zhang2008,Braithwaite2009,Hu2011} and chemical vapor transport\cite{Patel2009,Hara2010}. A breakthrough came with the use of a eutectic mix of KCl and AlCl$_3$ salts with low melting temperature to obtain FeSe directly in its tetragonal phase below 450$^\circ$C \cite{Chareev2013,Boehmer2013,Boehmer2016II} . This significantly improved the crystal quality, as shown by an up to ten-fold increase in residual resistivity ratio with respect to previously available samples \cite{Kasahara2014}. A technique for flux-free growth of large crystals with preferential orientation is described in Ref. \cite{Ma2014}. 

This topical review focuses on recent work using this 'new-generation' single crystals of FeSe and only occasionally refers to earlier results.

\subsection{Outline}
Early studies on iron-chalcogenide superconductors have been reviewed in Refs. \cite{Wu2009,Mizuguchi2010II,Deguchi2012,Singh2012,Malavasi2012}. 
Recently, the electronic structure of FeSe-related compounds has been reviewed in Refs. \cite{LiuJPCM2015} (with a focus on the single-layer materials) and \cite{Pustovit2016,Coldea_2017_review}. The monolayer thin films are furthermore subject of the reviews \cite{Sadovskii2016, Wang2017,Huang2017}. 

This topical review discusses nematicity, magnetism and superconductivity and the interplay of these phases in bulk FeSe. The orthorhombic/nematic phase of FeSe at ambient pressure, which has received a lot of attention, is described in section \ref{sec:nematicity}. Magnetic fluctuations and the pressure-induced magnetic order are reviewed in section \ref{sec:magnetism} and superconductivity is discussed in section \ref{sec:superconductivity}. Chemically substituted FeSe is briefly discussed in section \ref{sec:substitution}. The observed unusual interplay between structure, magnetism and superconductivity in FeSe has inspired a variety of theories, which will be reviewed in section \ref{sec:theory}. Furthermore, the theory of superconducting pairing in FeSe will be discussed in section \ref{sec:theory}. A summary and a comparison between FeSe and the archetypal 122-type iron-based systems will conclude this topical review.

\section{Nematicity}\label{sec:nematicity}

FeSe undergoes a tetragonal-to-orthorhombic phase transition similarly to other iron-based systems on decreasing temperature, but, unlike many other iron-based materials, this transition is not followed by magnetic order. Therefore, the orthorhombic phase of FeSe has received great attention as an opportunity to study the purely nematic phase in an iron-based superconductor over a wide temperature range. 
However, it is still an open question whether nematicity in FeSe is of the same origin as, e.g., in 122-type systems.

\subsection{Orthorhombic lattice distortion}

\begin{figure}
	\includegraphics[width=\textwidth]{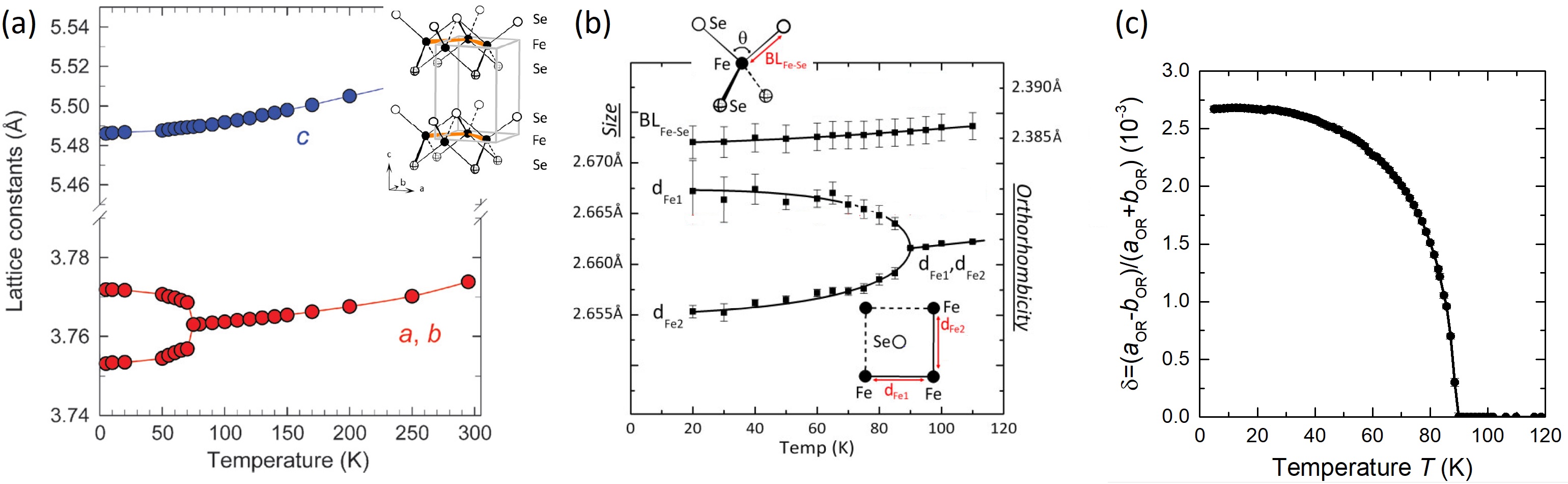}
	\caption{Temperature-dependent structural parameters of FeSe. (a) Temperature-dependent lattice parameters $a$, $b$ and $c$ of FeSe showing the tetragonal-to-orthorhombic transition in an early measurement using polycrystalline FeSe. (b) Fe-Se bond length and Fe-Fe distances as a function of temperature for a polycrystalline sample. Only the Fe-Fe distances are affected by the structural transition. (c) Temperature dependence of the orthorhombic distortion $\delta=(a-b)/(a+b)$ from high-energy x-ray diffraction on a single crystal. (a) Reproduced from Ref. \cite{Margadonna2008} with permission of The Royal Society of Chemistry. (b) Reproduced with permission from \cite{McQueen2009}, copyright 2009 American Physical Society. (c) Adapted from Ref. \cite{Kothapalli2016}, licensed under a Creative Commons Attribution 4.0 International License.}
	\label{fig:FeSe_orthorhombicity}
\end{figure}

\begin{figure}
	\includegraphics[width=\textwidth]{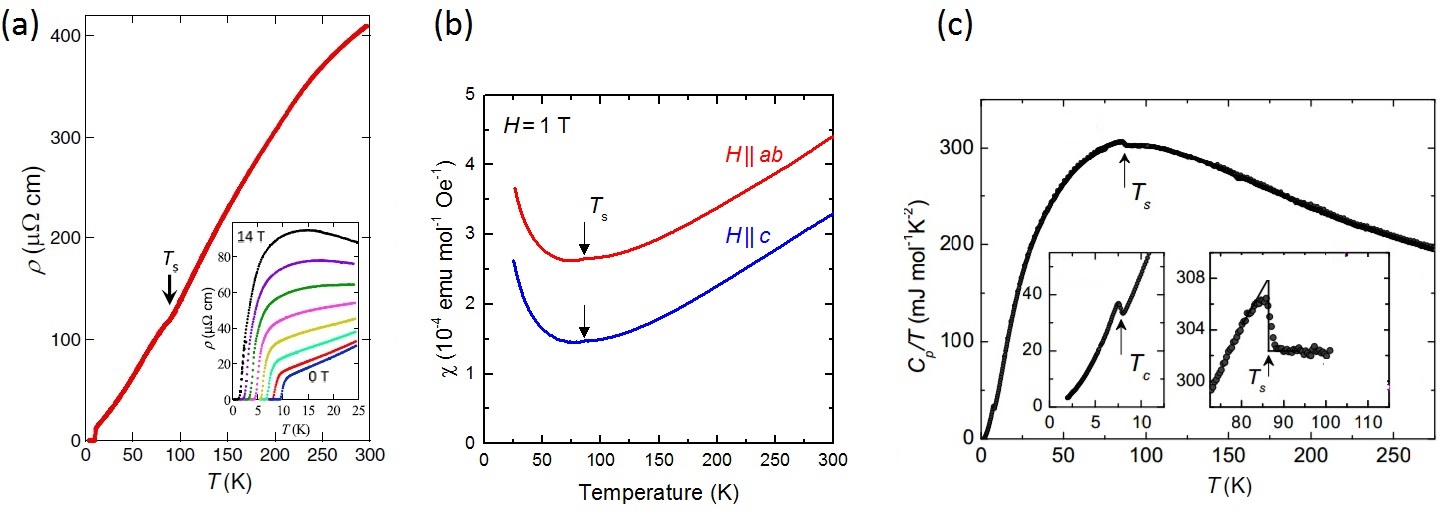}
	\caption{Signatures of the structural transition in standard transport and thermodynamic probes. (a) Resistivity of a high-quality FeSe single crystal with high residual resistivity ratio. The magneto-resistance reaches large values at low temperatures (inset). (b) Uniform magnetic susceptibility of an FeSe single crystal for revealing a subtle downward kink at $T_\mathrm{s}$ for both in-plane and out-of-plane external magnetic field\cite{Boehmerunpublished}. (c) Heat capacity divided by temperature of an FeSe single crystal. Insets show magnification of the data close to $T_\mathrm{s}$ and $T_\mathrm{c}$ revealing clear second-order phase transitions. (a) Reproduced from Ref. \cite{Kasahara2014}. (c) Reproduced with permission from Ref. \cite{Boehmer2015}, copyright 2015 American Physical Society.}
	\label{fig:FeSe_characterization}
\end{figure}

The tetragonal-to-orthorhombic structural transition at $T_\mathrm{s}$ in FeSe was observed already at the time of the discovery of superconductivity in the material \cite{Hsu2008,Margadonna2008} (Fig. \ref{fig:FeSe_orthorhombicity} a). The structural deformation is analogous to the ubiquitous tetragonal-to-orthorhombic distortion in many 1111-, 122-, and 111-type iron-based materials. The nearest-neighbor Fe-Fe distances $a$ and $b$ -- directed along the in-plane diagonal of the tetragonal unit cell -- become inequivalent, resulting in a doubling and rotation of the tetragonal unit cell in the basal plane. The Fe-Se bond-lengths are not affected (Fig. \ref{fig:FeSe_orthorhombicity} b).

The temperature dependence of the orthorhombic distortion \mbox{$\delta=(a-b)/(a+b)$} indicates a continuous (second-order) phase transition at $T_\mathrm{s}=90$ K. At low temperatures, $\delta\approx2.7\times10^{-3}$ \cite{Kothapalli2016} (see Fig. \ref{fig:FeSe_orthorhombicity}c). This value is very similar to the orthorhombic distortion of BaFe$_2$As$_2$ ($\delta(T=0)=4\times10^{-3}$ \cite{Avci2011}) when accounting for the different transition temperature, $T_\mathrm{s}\approx140$ K. In fact, the ratio $\delta(T=0)/T_\mathrm{s}$ is practically identical in the two systems. Intriguingly, the onset of superconductivity does not measurably change the orthorhombic distortion (Fig. \ref{fig:FeSe_orthorhombicity} c), in strong contrast to 122-type iron-based superconductors\cite{Boehmer2013}.

The structural transition of FeSe is also visible in standard transport and thermodynamic probes (Fig. \ref{fig:FeSe_characterization}). Resistivity shows a slight upward kink on decreasing temperature through $T_\mathrm{s}$. This anomaly is more pronounced in single crystals and was sometimes overlooked in polycrystalline material. 
The uniform magnetic susceptibility has a slight downward kink at $T_\mathrm{s}$. The specific heat shows a clear second-order type jump at $T_\mathrm{s}$ of $\Delta C_p/T_\mathrm{s}\approx 6$ mJ mol$^{-1}$ K$^{-2}$\cite{Boehmer2015,Karlsson2015}, as expected for a continuous phase transition. Interestingly, the size of the specific-heat anomaly is very similar to the low-temperature Sommerfeld coefficient of $\sim 6-7$ mJ mol$^{-1}$, suggesting an electronic phase transition\cite{Karlsson2015}.

 \subsection{Electronic in-plane anisotropy} 
 
 \begin{figure}
	\includegraphics[width=\textwidth]{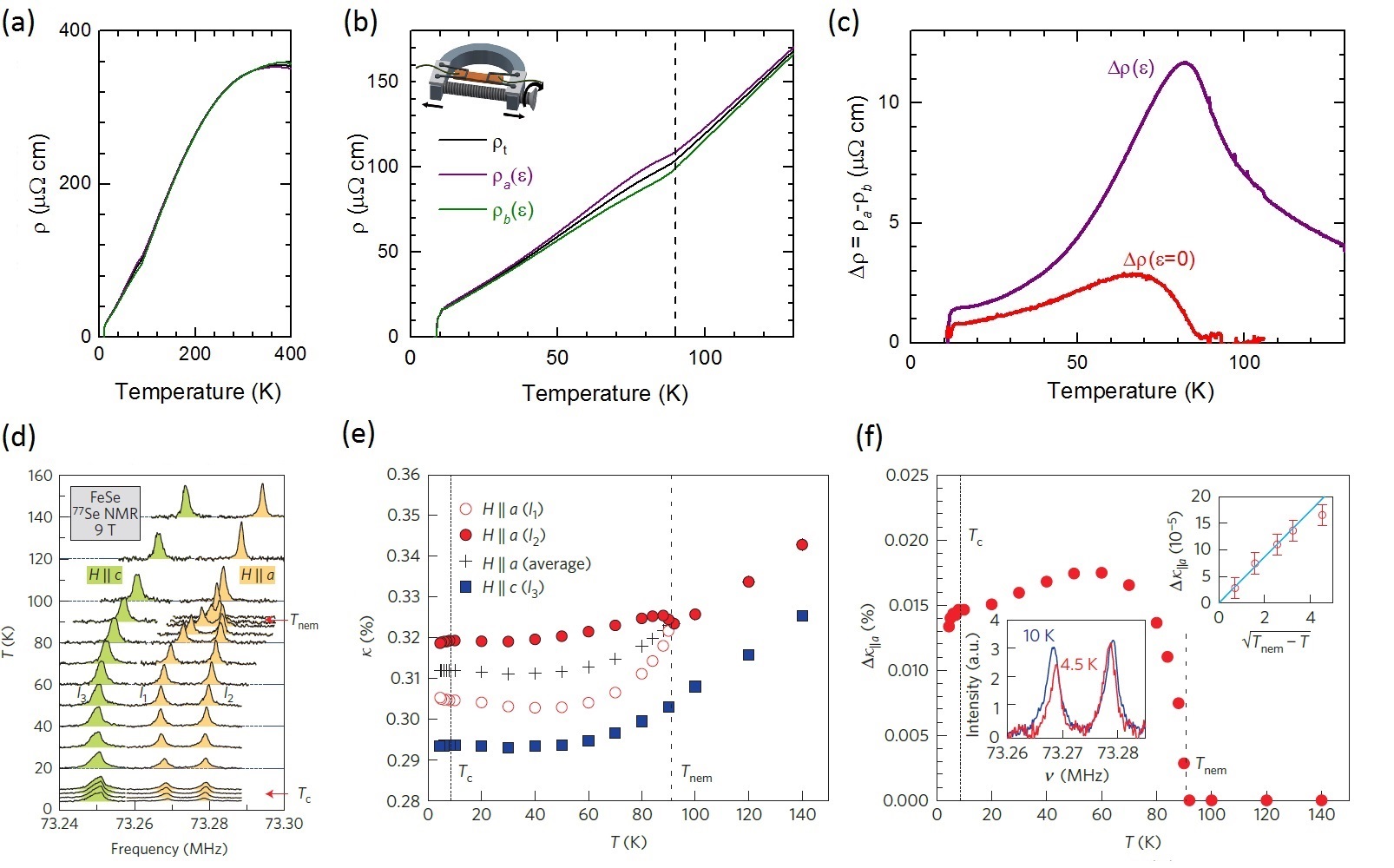}
	\caption{In-plane anisotropy of electronic properties of FeSe. (a), (b) In-plane resistivity of twinned ($\rho_t$) and detwinned ($\rho_a$, $\rho_b$) FeSe. Samples were detwinned by applying tensile strain, $\varepsilon$, via a 'horseshoe device' [inset in (b)], and $\rho_b$ was calculated from the two measurements\cite{Tanatar2016}. (c) The measured resistivity anisotropy $\Delta\rho=\rho_a-\rho_b$ at finite applied strain (purple line) consists of an 'intrinsic' $a$-$b$ anisotropy of \mbox{$\leq3$ $\mu\Omega$cm} (red line) and a larger strain-induced contribution \cite{Tanatar2016}. (d) NMR spectra of an FeSe single crystal, revealing a significant different NMR spectral shift for the two types of orthorhombic domains below $T_\mathrm{s}$ for a given field direction {\red as a line splitting}. (e) Temperature dependent NMR spectral shift. (f) The difference of the two in-plane NMR lines, showing an order-parameter-like temperature dependence close to $T_\mathrm{s}$ and a slight decrease below $T_\mathrm{c}$ \cite{Baek2015}. (a)-(c) Reproduced with permission from Ref. \cite{Tanatar2016}, copyright 2016 American Physical Society. (d)-(f) Reprinted by permission from Macmillan Publishers Ltd: Nature Materials, Ref. \cite{Baek2015}, copyright 2015.}
	\label{fig:FeSe_electronicanisotropy}
\end{figure}
 
The observation of a large in-plane resistivity anisotropy in detwinned samples of Co-doped BaFe$_2$As$_2$\cite{Tanatar2010II,Chu2010} was a starting point for the investigation of electronic nematicity in the iron-based systems.
FeSe single crystals are easily damaged by uniaxial force, which makes the mechanical detwinning necessary for a measurement of this in-plane anisotropy difficult. Only measurements {\red of the resistivity anisotropy} under tensile uniaxial stress in a 'horseshoe device' have succeeded so far\cite{Tanatar2016} (Fig. \ref{fig:FeSe_electronicanisotropy} (a),(b)) . A small resistivity anisotropy with $\rho_a>\rho_b$ was found. The sign of $\rho_a-\rho_b$ is opposite to (electron-doped) BaFe$_2$As$_2$. Furthermore, a significant fraction of the observed anisotropy was shown to be induced by the applied uniaxial stress. Subtraction of this elastoresistive effect from the measured $\rho_a-\rho_b$ indicates an anisotropy of $\rho_a-\rho_b \leq 3$ $\mu\Omega$cm in the limit of zero external force, which is only about 4\% of the in-plane average resistivity; a small value compared to Ba(Fe$_{1-x}$Co$_x$)$_2$As$_2$\cite{Ishida2013} (Fig. \ref{fig:FeSe_electronicanisotropy} (c)). 

Nuclear magnetic resonance (NMR) as a local probe allows to distinctly study the two types of orthorhombic domains\cite{Fu2012} without the need for detwinning. $^{77}$Se is an NMR active nucleus and $^{77}$Se NMR has played an important role in the study of FeSe \cite{Kotegawa2008III,Imai2009,Masaki2009,Boehmer2015,Baek2015,Baek2016,Wang2016}. Under a given direction of in-plane applied magnetic field, the two types of orthorhombic domains show distinctly different NMR spectral shifts, $\mathcal K$. This difference, $\Delta  \mathcal{K}$ arises because the magnetic field is directed along the $a$-axis of one type and along the $b$-axis of the other type of domains \cite{Boehmer2015,Baek2015,Wang2016} [Fig. \ref{fig:FeSe_electronicanisotropy} (d-f)]. The value of the NMR spectral shift (Knight shift) is determined by the local magnetic susceptibility at the Se site and the hyperfine coupling constant. In Ref. \cite{Baek2015}, it was argued that the observed significant line splitting cannot be a 'simple' consequence changing atomic distances following the small orthorhombic lattice distortion. Moreover, $\Delta \mathcal{K}$ decreases below the superconducting transition, in contrast to the structural distortion. For these reasons, the anisotropy $\Delta  \mathcal{K}$ has been associated with an electronic nematic order parameter\cite{Baek2015}. From the absence of a strong response of spin fluctuations to the nematic transition, it was argued that nematicity in FeSe is likely of orbital origin\cite{Baek2015}. In Ref. \cite{Boehmer2015}, it was shown that the signature of spin fluctuations in NMR sets in at $T_\mathrm{s}$, but is absent at higher temperatures. From this observation, the driving force of the structural transition in FeSe was also argued to be non-magnetic. However, later inelastic neutron scattering showed the presence of magnetic fluctuations at all temperatures (see section \ref{sec:magnetism}), as well as a feedback effect between spin-fluctuations and orthorhombic distortion\cite{Wang2015,WangLee2015NatPhys_FeSe-Para-Nematicity}, indicating that nematicity in FeSe may have an intimate relation with magnetism after all.

The electronic in-plane anisotropy of FeSe was also revealed and studied using optical methods, namely in the optical {\red reflectivity in the mid-infrared spectral range\cite{Chinotti2017}, in the optical bireflectance \cite{Tanatar2016}, and in polarized ultrafast spectroscopy in a pump-probe experiment \cite{Luo2017}. Recently, the in-plane anisotropy in the nematic state and associated fluctuations were studied using resistance, uniform magnetic susceptibility and magnetostriction measurements of single crystals strained via a substrate \cite{He2017}}. 

\subsection{Nematic susceptibility}

\begin{figure}
	\includegraphics[width=\textwidth]{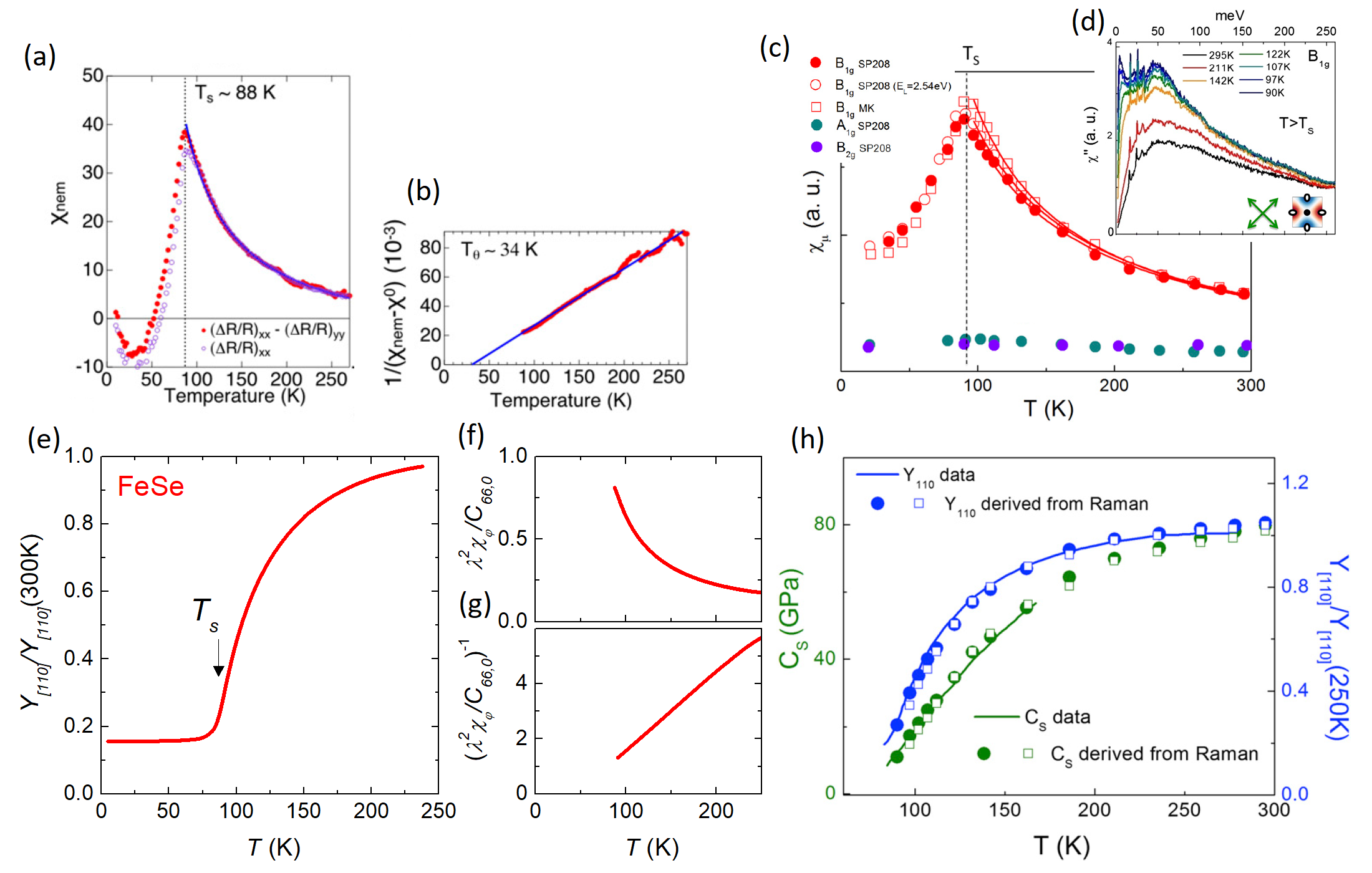}
	\caption{Nematic susceptibility of FeSe. (a) Nematic susceptibility $\chi_{\mathrm{nem}}$ of FeSe as measured by the elastoresistivity coefficient $m_{66}$ and (b), a Curie-Weiss plot of the same quantity\cite{Hosoi2016}. (c) Charge nematic susceptibility from electronic Raman scattering, diverging on approaching $T_\mathrm{s}$ \cite{Massat2016}. (d) shows the a selection of the corresponding Raman spectra. (e) Normalized Young's modulus $Y_{[110]}$ of FeSe, dominated by the elastic shear modulus $C_s$, obtained by three-point bending\cite{Boehmer2016}. Assuming the existence of a distinct nematic order parameter, the nematic susceptibility can be inferred from the shear modulus as $C_{s}/C_{s,0}=1-\lambda^2\chi_{\mathrm{nem}}$ ( $C_{s,0}$ is bare value of the shear modulus and $\lambda$ is a coupling constant), and is shown in (f), (g). (h) Under the same assumption, the charge nematic susceptibility in (c) is perfectly consistent with $Y_{[110]}$ and $C_s$ \cite{Massat2016}. (a), (b) Reproduced from Ref. \cite{Hosoi2016}. (c), (d), (h) Reproduced from Ref. \cite{Massat2016}. (e)-(g) Adapted from Ref. \cite{Boehmer2016}. }
	\label{fig:7}
\end{figure}

The nematic susceptibility has been one of the key properties in the study of iron-based systems. This quantity describes the tendency of the material to develop $a$-$b$ anisotropy under uniaxial strain. It can be determined as the in-plane anisotropy of a physical quantity, e.g., resistivity, that is induced by elastic deformation of the sample \cite{Chu2012}. 
The nematic susceptibility defined in this way tends to a divergence at a temperature $T_0<T_\mathrm{s}$, which would be the nematic transition in the absence of the coupling to the crystal lattice. The ``real'' nematic susceptibility, however, is renormalized by this coupling to the lattice so that it diverges as expected when nematic order sets in at $T_\mathrm{s}$\cite{Chu2012,Boehmer2016}.

The elastoresistivity coefficient, $2m_{66}=\frac{1}{\rho}\frac{d\left(\Delta\rho\right)}{d\left(\varepsilon_a - \varepsilon_b\right)}$, describes the relative change of the resistance anisotropy $\Delta\rho=\rho_a-\rho_b$ with respect to orthorhombic strain and is frequently used as a measure of the nematic susceptibility\cite{Chu2012,Kuo2013,Kuo2016}. It is found to diverge upon approaching $T_\mathrm{s}$ in FeSe \cite{Watson2015,Hosoi2016,Tanatar2016} [Fig. \ref{fig:7}(a),(b)], consistent with an electronically driven nematic transition. The magnitude of $2m_{66}$ is comparable to BaFe$_2$As$_2$ in two reports\cite{Tanatar2016,Hosoi2016} but its sign is opposite to BaFe$_2$As$_2$. The temperature dependence of $2m_{66}$ follows approximately a Curie-Weiss law $\sim\frac{1}{T-T_0}$ and the inferred Curie temperature $T_0$ ranges from 34 K to 83 K in the different reports \cite{Watson2015,Hosoi2016,Tanatar2016}. 

The elastic shear modulus related to the structural transition of FeSe, $C_s$, softens (decreases) significantly on approaching $T_\mathrm{s}$ from above \cite{Zvyagina2013,Boehmer2015}, Fig. \ref{fig:7}(e). This is expected for a second order structural transition regardless of its origin. In the case of an electronic nematic phase transition, the bare shear modulus $C_\mathrm{s,0}$ is renormalized by the nematic susceptibility $\chi_\mathrm{nem}$, according to $C_\mathrm{s}=C_\mathrm{s,0}-\lambda^2\chi_\mathrm{nem}$\cite{Fernandes2010,Boehmer2014}. The inferred $\chi_\mathrm{nem}$ follows a perfect Curie-Weiss law and is consistent with the elastoresistivity measurements (Fig. \ref{fig:7}(f),(g)).

The electronic charge susceptibility in the symmetry channel  that is associated with the nematic transition can be obtained from electronic Raman scattering \cite{Gallais2014,Gallais2016} (Fig. \ref{fig:7}(d)). Notably, the thus inferred nematic susceptibility of FeSe, $\chi_{B1g}$, also shows a clear divergence \cite{Massat2016} (Fig. \ref{fig:7}(c)). The $\chi_{B1g}$-data can be used to explain the elastic shear modulus via $C_\mathrm{s}=C_\mathrm{s,0}-\lambda^2\chi_\mathrm{nem}$. The experimental data\cite{Zvyagina2013,Boehmer2015,Massat2016} are in perfect agreement (Fig. \ref{fig:7}(h)), which has been interpreted as evidence for charge-induced nematicity in FeSe\cite{Massat2016}.

\subsection{Electronic structure}

\begin{figure}
	\includegraphics[width=\textwidth]{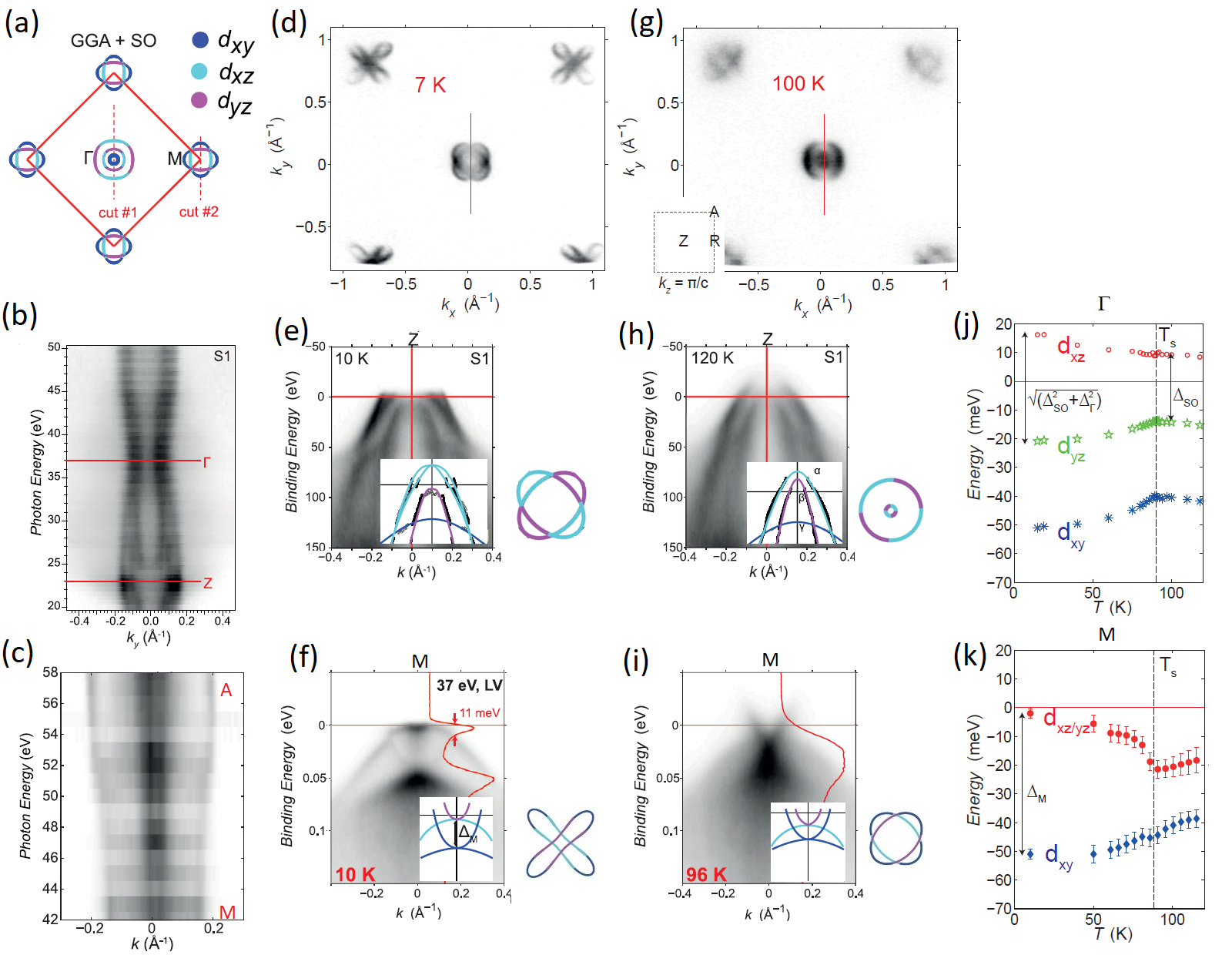}
	\caption{High-resolution ARPES data on twinned FeSe\cite{Watson2015,Watson2016,Coldea_2017_review}. (a) Fermi surfaces according to DFT calculations predict hole cylinders around the zone center and electron cylinders in the zone corners with orbital contents as indicated. (b), (c) Measured $k_z$ dispersion of the Fermi surface cylinders {\red close to the zone center and the zone corner, respectively. (d)-(f) Fermi surface map and cuts across the $Z$ and $M$ points above $T_s$. A schematic representation of the dispersions and the Fermi surfaces are indicated. (g)-(i) Fermi surface map and cuts across the $Z$ and $M$ points below $T_s$. A schematic representation of the dispersions and the Fermi surfaces are indicated. Two ellipses corresponding to two types of orthorhombic domains are visible.} (j) Temperature dependence of the position of $d_{xz}$, $d_{yz}$ and $d_{xy}$ bands at the $\Gamma$ point. The splitting between the bands of $d_{xz}$ and $d_{yz}$ character above $T_\mathrm{s}$ arises from spin-orbit coupling. (k) Temperature dependence of the position of $d_{xz}$, $d_{yz}$ and $d_{xy}$ bands at the $M$ point. No splitting between $d_{xz}$ and $d_{yz}$ is resolved. (a), (b), (e), (h) Reproduced with permission from Ref. \cite{Watson2015}, copyright 2015 American Physical Society. (c), (d), (f), (g), (i), (k) Reproduced from Ref. \cite{Watson2016}, Creative Commons Attribution 3.0 licence. (j) Reproduced with permission from Ref. \cite{Coldea_2017_review}.}
	\label{fig:FeSe_ARPES}
\end{figure}

\begin{figure}
	\includegraphics[width=\textwidth]{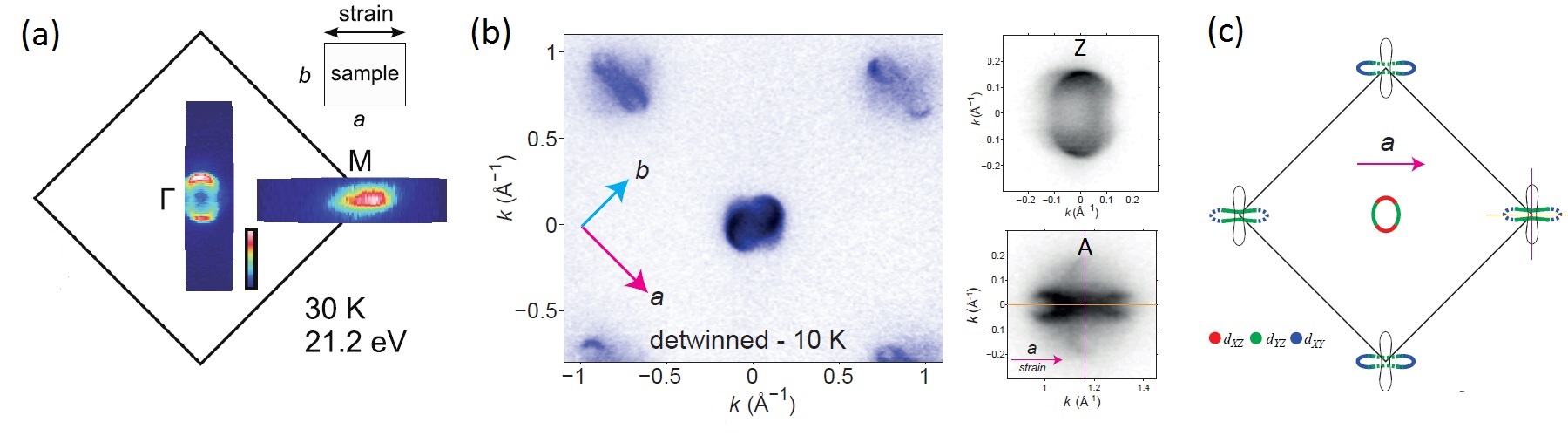}
	\caption{ARPES data on detwinned FeSe (a) Low-temperature Fermi surface map obtained by ARPES on detwinned crystals of FeSe, showing hole and electron pockets elongated in opposite directions. The schematic drawing illustrates the orientation of the sample axes with respect to the applied strain. (b) High-resolution Fermi surface map of $\sim80\%$ detwinned FeSe and details around the hole and electron pockets. Most surprisingly, a selection rule appears to prevent observation of one of the two ``peanut''-shaped pockets in the zone corner. (c) Schematic illustration of the Fermi surface of the majority domain (colored lines, indicating orbital content) and contribution of the minority domain (thin black line). (a) Reproduced with permission from Ref. \cite{Suzuki2015}, copyright 2015 American Physical Society. (b), (c) Reproduced from Ref. \cite{Watson2017II}, Creative Commons Attribution 3.0 licence.}
	\label{fig:FeSe_ARPES_detwinned}
\end{figure}

The electronic structure of FeSe is complex and has been studied in detail via magnetotransport, quantum oscillations, ARPES and STM. The magneto-resistance of FeSe increases sharply below $T_\mathrm{s}$\cite{Huynh2014,Knoener2015}. Detailed analysis of magneto-transport data in multiband models indicates the reduction of total carrier density and the emergence of a small number of high-mobility carriers\cite{Huynh2014,Sun2016II,Sun2016III,Terashima2016II,Watson2015II,Ovchenkov2017} in the orthorhombic phase.

A number of ARPES studies have investigated the electronic structure of FeSe in great detail and have recently been reviewed in \cite{LiuJPCM2015,Pustovit2016,Coldea_2017_review}. Here, a short summary is presented, with a focus on the changes of the electronic structure related to the nematic transition (Fig. \ref{fig:FeSe_ARPES}).  

The electronic structure of FeSe is broadly similar to other iron-based superconductors and the Fermi surface consists of quasi two-dimensional hole cylinders around the Brillouin zone center and electron cylinders around Brillouin zone corners, seen in ARPES  (Fig. \ref{fig:FeSe_ARPES} (b)-(e)). The size of the observed Fermi surfaces is significantly reduced with respect to band-structure calculations in the tetragonal state (Fig. \ref{fig:FeSe_ARPES}(a)), especially at low temperatures\cite{Rhodes2017,Kushnirenko_controversial}. Quantum oscillation measurements \cite{Terashima2014,Watson2015II,Audouard2015EPL_Hc2} are consistent with these results. 

Three bands are close to the Fermi level in the Brillouin zone center (Fig. \ref{fig:FeSe_ARPES}(f)). The $\alpha$-band crosses the Fermi level and forms a hole-type Fermi surface cylinder with a mild $k_z$ dispersion (Fig. \ref{fig:FeSe_ARPES}(b)). The $\beta$-band crosses the Fermi level only close to the Z point and thus forms a three-dimensional pocket \cite{Watson2015,Zhang2015_PRB}. The $\alpha$ and $\beta$ bands are of $d_{xz}/d_{yz}$ character and are split by 20 meV at $\Gamma$, via spin-orbit interaction\cite{BorisenkoSO,Watson2016II}. Additionally, there is a $d_{xy}$-type $\gamma$ band lying entirely below the Fermi energy. It displays a particularly strong mass renormalization of $\sim9$.

Close to the Brillouin zone center, there are only subtle changes of the band structure associated with the structural transition. On entering the orthorhombic phase, the splitting between $\alpha$ and $\beta$ bands increases (Fig. \ref{fig:FeSe_ARPES}(j)) so that the $\beta$ band is pushed entirely below the Fermi energy. In addition, there is an elliptical distortion of the remaining hole Fermi surface. The superposition of two types of orthorhombic domains leads to the observed crossed ellipses at the center of the Brillouin zone, Fig. \ref{fig:FeSe_ARPES} (e),(g) \cite{Watson2015}, though this scenario was questioned in Ref. \cite{Pustovit2016}. Measurements on detwinned crystals (see Fig. \ref{fig:FeSe_ARPES_detwinned}) have indicated that orbital energy of the $d_{xz}$ orbital is higher than the energy related to the $d_{yz}$ orbital, $E_{xz}>E_{yz}$, in the Brillouin-zone center\cite{Suzuki2015}. The difference was estimated to $\sim10$meV \cite{Suzuki2015} (15 meV in Refs. \cite{Watson2016,Fedorov2016}).
 
The situation at the Brillouin-zone corner is more complex. Two elliptical, quasi two-dimensional electron pockets are observed at the M-point (Fig. \ref{fig:FeSe_ARPES} (d), (e), (h), (i)). A large splitting of 50 meV between two bands at the M point at low temperatures was initially associated with the difference between orbital energies $E_{xz}$ and $E_{yz}$ in the orthorhombic phase \cite{Shimojima2014,Nakayama2014,Watson2015}. This appears consistent with recent data presented in Ref. \cite{Fanfarillo2016}. This large value of 50 meV, which cannot be reproduced by band-structure calculations of orthorhombic FeSe, has been taken as evidence for the presence of electronic nematic order in the material (e.g., Ref. \cite{Shimojima2014}). The measurements on detwinned crystals indicate that $E_{xz}<E_{yz}$ \cite{Shimojima2014,Suzuki2015}, i.e., the sign of orbital order at the M point is opposite to the $\Gamma$-point \cite{Suzuki2015}. 
 
New high-resolution data on twinned and detwinned samples along with a reinterpretation of the band splitting at the M point have been reported recently\cite{Watson2016,Fedorov2016,Watson2017II}. 
According to this interpretation, the band splitting of 50 meV at the M point does not arise from nematic order and is present at high temperatures already (Fig. \ref{fig:FeSe_ARPES} (k)). On the other hand, a splitting between $d_{xz}$ and $d_{yz}$ associated with nematic order of only $\sim 10$ meV at the lowest temperature was suggested \cite{Fedorov2016}. Notably, it was pointed out in Ref. \cite{Fedorov2016} that this splitting of 10 meV at the M point is entirely consistent with band structure calculations in the orthorhombic state and no additional electronic nematic order is required to explain it. Only the orbital anisotropy at the Brillouin-zone center is not predicted by these calculations. The implication is that the electronic nematic order is of a type that affects only the orbital energies close to the center of the Brillouin zone. A candidate, unidirectional nematic bond ordering, has been proposed in Ref. \cite{Watson2016}. 

The recent high-resolution ARPES measurements on detwinned crystals \cite{Watson2017II} (Fig. \ref{fig:FeSe_ARPES_detwinned}(b),(c)) reveal that, as an unexpected dramatic consequence of nematicity, only one of the two crossed ``peanut-shaped'' electron pockets can be observed when  a single orthorhombic domain is studied. It was suggested that this might be due to a selection rule specific to ARPES, whose origin is unknown. Similarly, in recent Boguliubov quasiparticle interference experiments, only one electron pockets was observed, which was ascribed to a significantly reduced coherence of the second pocket\cite{Sprau2017}. The relation between this ``one-peanut'' observation and nematic order is a fascinating open question.

\section{Magnetism}\label{sec:magnetism}

At ambient pressure, bulk FeSe does not order magnetically, but significant magnetic fluctuations with a complex temperature and momentum dependence are observed. The application of hydrostatic pressure induces a dome of magnetic order with a maximum transition temperature $T_\mathrm{N}\approx45$ K.  

\subsection{Magnetic fluctuations at ambient pressure}

\begin{figure}
	\includegraphics[width=\textwidth]{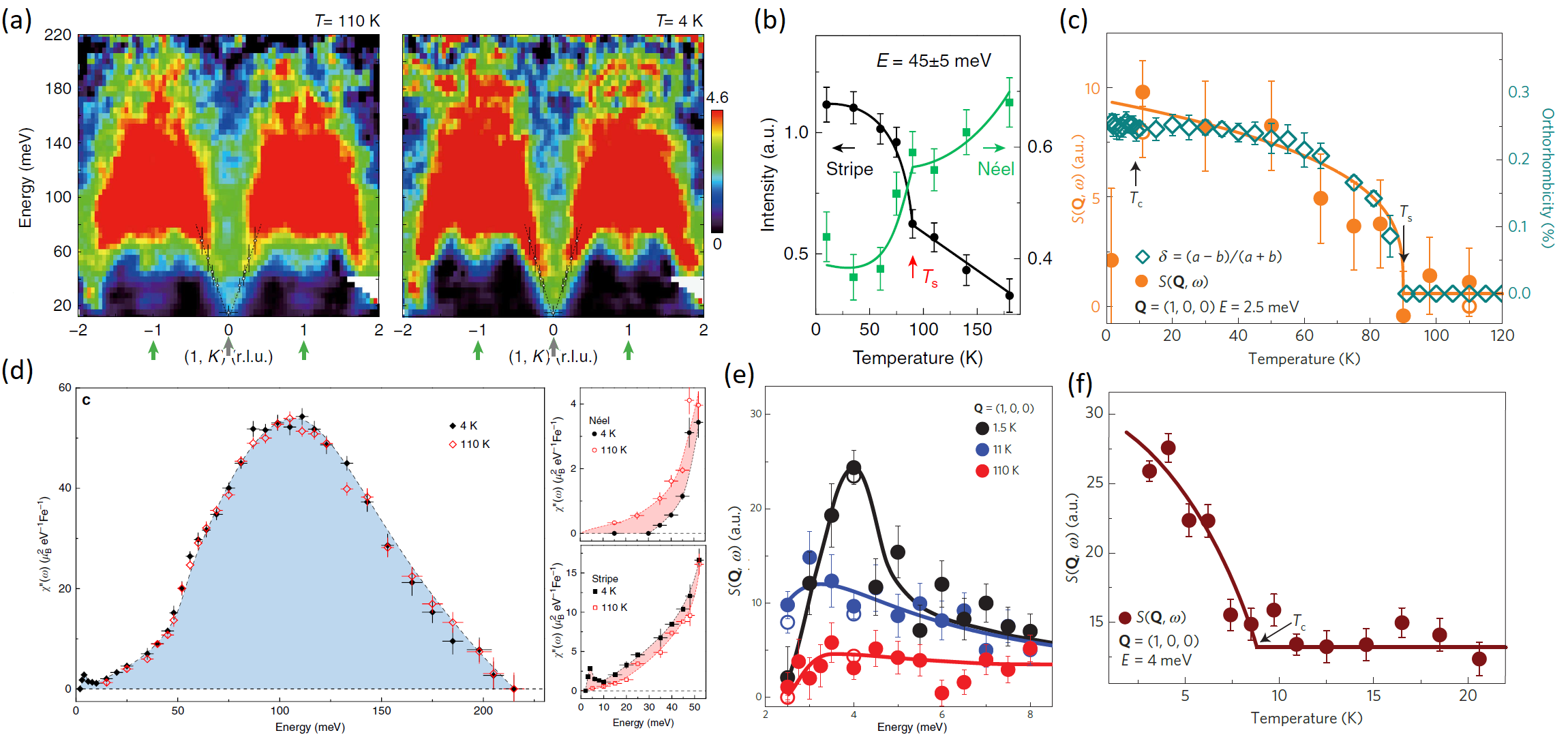}
	\caption{Inelastic neutron scattering results from coaligned sets of FeSe single crystals \cite{Wang2015,Wang2015GS}. (a) Dispersion of spin fluctuations in FeSe at $T=110$ K and 4 K, showing intensity both around the stripe-type wave vector and the N\'eel-type wave vector [$(1,0)$ and $(1,1)$, respectively, in this notation]. (b) Temperature dependence of the intensity around the stripe-type and N\'eel-type wave vectors. (c) Temperature dependence of the low-energy dynamic spin correlation $S(Q,\omega)$, around the stripe-type wave vector (yellow circles), which correlates with the orthorhombic structural distortion (teal diamonds). (d) Energy dependence of the dynamic susceptibility at 4 and 110 K, corresponding to the data in (a) and the energy dependence of the N\'eel-type and stripe-type fluctuations. (e) Energy dependence of the dynamic spin correlation at low temperatures, revealing a superconducting resonance at the stripe-type wave vector. (f) Temperature dependence of the low-energy dynamic spin correlation around the stripe-type wave vector enhanced by the resonance below $T_\mathrm{c}$. (a),(b),(d) Reproduced from Ref. \cite{Wang2015GS}, Creative Commons Attribution 4.0 International License. (c),(e),(f) Reprinted by permission from Macmillan Publishers Ltd: Nature Materials, Ref. \cite{Wang2015}, copyright 2015.}
	\label{fig:8}
\end{figure}

An early review\cite{Kotegawa2012} discussed magnetic excitations in FeSe from NMR and inelastic neutron scattering experiments.  The early NMR measurements on polycrystalline FeSe\cite{Imai2009} show an enhancement of the spin-lattice relaxation rate {\red divided by temperature} $1/T_1T$ arising from an increase in low-energy spin fluctuations around $T^*\approx100$ K. Subsequent measurements on single crystals showed that the onset of this enhancement occurs very close to $T_\mathrm{s}$\cite{Boehmer2015} at ambient pressure, suggesting a coupling of magnetic fluctuations and nematic order. 

Recently, the magnetic fluctuations of FeSe at ambient pressure have been studied in detail using inelastic neutron scattering\cite{Rahn2015,Wang2015,Wang2015GS,Ma_2017} revealing their complex dependence on temperature and wave vector (Fig. \ref{fig:8}) . Magnetic fluctuations are present both around the stripe-type wave vector and around the N\'eel-type wave vector over a broad energy range\cite{Wang2015GS} (Fig. \ref{fig:8} (a)). They were found to have a smaller bandwidth than in 122-type systems. A large corresponding total fluctuating moment of \mbox{$\sim 5.1$ $\mu_\mathrm{B}^2$/Fe} (effective spin of $S\sim0.74$) was estimated \cite{Wang2015GS}. At very low energies, magnetic fluctuations have negligible spectral weight for $T>T_\mathrm{s}$ but acquire spectral weight below $T_\mathrm{s}$\cite{Wang2015} (Fig. \ref{fig:8}(c)), consistent with the NMR data \cite{Vaknin_16}. On decreasing temperature through $T_\mathrm{s}$, spectral weight is transferred from the checkerboard-type to the stripe-type fluctuations for  energies $\lesssim 60$ meV \cite{Wang2015GS} (Fig. \ref{fig:8}(b),(d)). {\red Very recently, electronic Raman spectra have been interpreted as arising from a frustrated spin-1 system with competition between stripe-type and N\'eel-type magnetic ordering vectors \cite{Baum2017}, consistent with the neutron results. On the other hand, recent Raman data on FeSe and FeSe$_{1-x}$S$_x$ in Ref. \cite{Zhang2017} have been taken as evidence for the formation of stripe-type quadrupolar order, i.e., a wave of quadrupole moment without
charge or spin modulation, in the nematic phase of FeSe.}

The presence of spin fluctuations in FeSe at ambient pressure indicates the proximity to a magnetically ordered state. However, the complexity of the fluctuation spectrum may indicate magnetic frustration, which is a possible explanation for the absence of magnetic ordering in FeSe at ambient pressure\cite{Glasbrenner2015}. The clear impact of the structural transition on the magnetic fluctuation spectrum demonstrates the coupling between magnetic and structural properties in FeSe \cite{Wang2015,Wang2015GS}.

In the superconducting state, a clear spin-resonance is observed around the stripe-type wave vector\cite{Wang2015} (Fig. \ref{fig:8}(e),(f)), consistent with a spin-fluctuation-mediated superconducting pairing mechanism. Polarized inelastic neutron scattering has revealed that spin-orbit coupling is important and that the low-energy magnetic fluctuations, including the superconducting resonance, are mainly $c$-axis polarized\cite{Ma_2017}.

\subsection{Pressure-temperature phase diagram}

\begin{figure}
	\includegraphics[width=\textwidth]{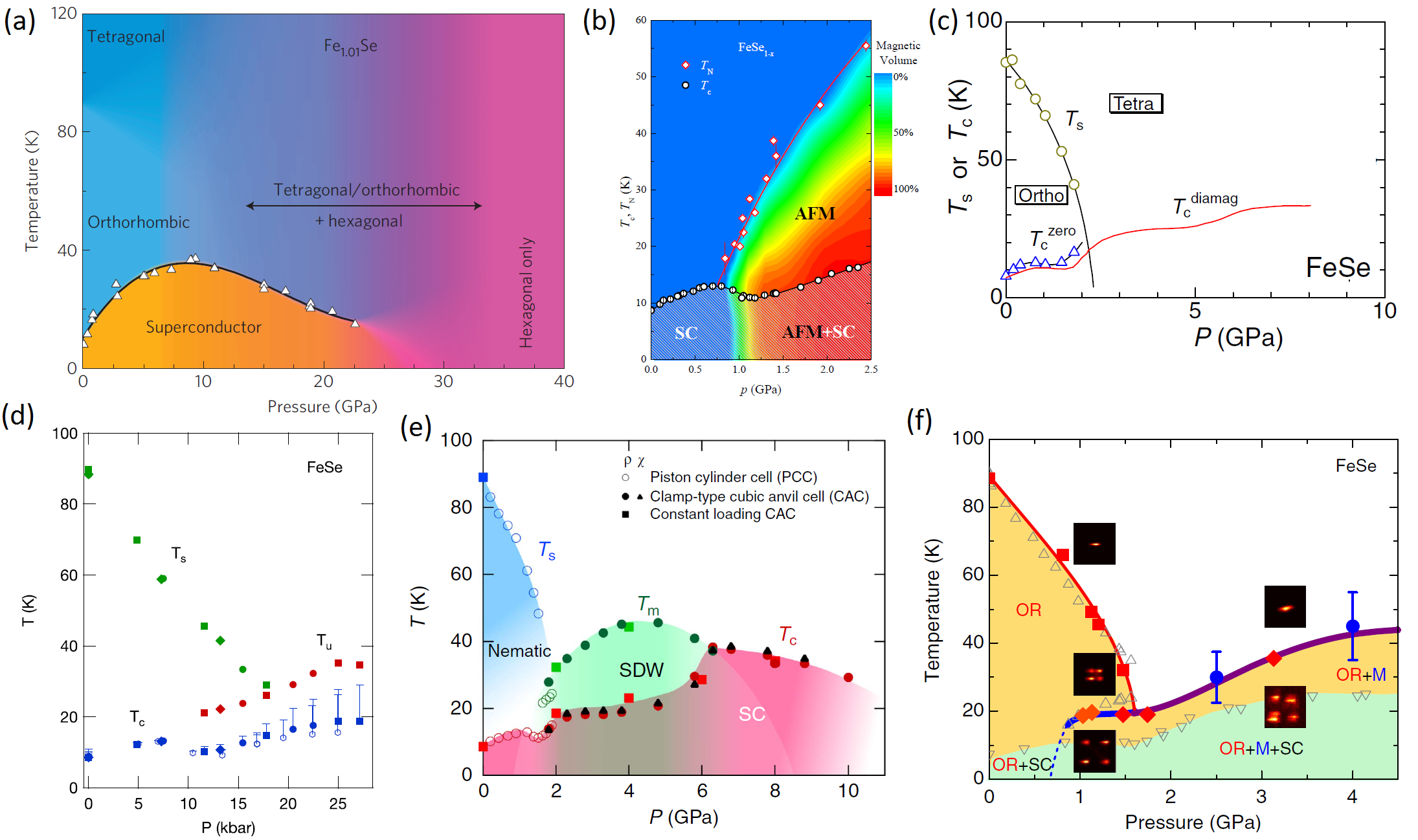}
	\caption{Experimental pressure-temperature phase diagrams of FeSe. (a) Early temperature-pressure phase diagram for polycrystalline FeSe, extending to high pressures. The huge enhancement of $T_\mathrm{c}$ under pressure and the transformation to a different crystallographic phase at high pressures is shown. (b) Phase diagram obtained from muSR measurements on the polycrystalline material, showing the emergence of magnetic order. (c) Phase diagram obtained from resistivity and magnetization measurements on single crystals, showing the suppression of $T_\mathrm{s}$ under pressure for the first time and detailing the ``three-step''-like enhancement of $T_\mathrm{c}$ under pressure. (d) Phase diagram {\red obtained from resistivity and susceptibility measurements on high-quality single crystals} revealing for the first time both structural and magnetic phase transitions in the same experiment. (e) Phase diagram obtained from resistivity measurements on high-quality single crystals over a large pressure range (see Fig. \ref{fig:10}), showing a dome-shaped region of magnetic order. (f) Phase diagram obtained using x-ray diffraction and M\"ossbauer spectroscopy on single crystals under pressure, revealing that structural and magnetic phase lines merge into a joint magneto-structural transition under pressure. (a) Reprinted by permission from Macmillan Publishers Ltd: Nature Materials, Ref. \cite{Medvedev2009}, copyright 2009. (b) Reproduced with permission from Ref. \cite{Bendele2012}, copyright 2012 American Physical Society. (c) Reproduced with permission from Ref. \cite{Miyoshi2014}, article copyrighted by JPS \copyright 2014, The Physical Society of Japan. (d) Reproduced from Ref. \cite{Terashima2015}, Creative Commons Attribution 4.0 License. (e) Reproduced from Ref. \cite{Sun_2016}, Creative Commons Attribution 4.0 International License. (f) Reproduced from Ref. \cite{Kothapalli2016}, Creative Commons Attribution 4.0 International License.}
	\label{fig:9}
\end{figure}

\begin{figure}
	\includegraphics[width=\textwidth]{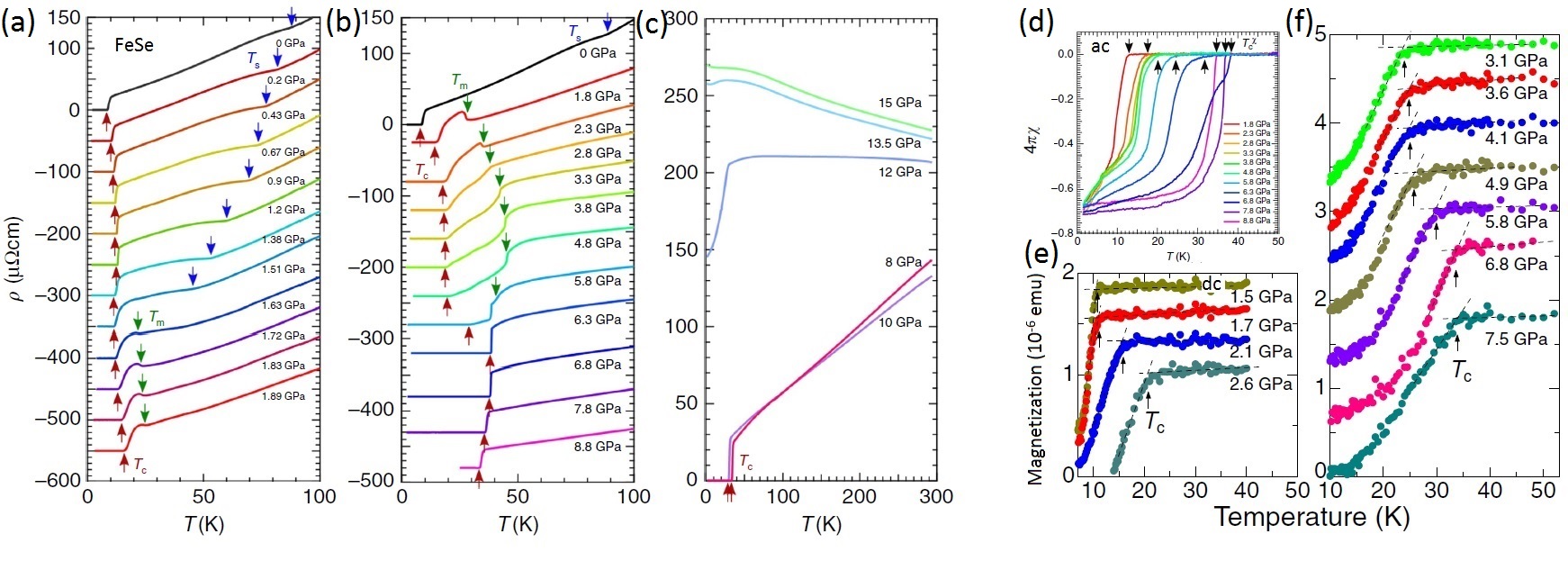}
	\caption{Resistance and magnetization data of single crystalline FeSe under pressure. (a)-(c) Resistivity data over a wide pressure range. A kink indicates the structural transition which is suppressed under pressure and an upturn (downturn at higher pressures) indicates the onset of magnetic order. At the highest pressures, the resistivity increases abruptly indicating the transition to the high-pressure orthorhombic phase. (d) AC susceptibility and (e), (f) DC magnetic susceptibility \cite{Miyoshi2014} showing a clear diamagnetic signal related to superconductivity over the whole pressure range. (a)-(d) Reproduced from Ref. \cite{Sun_2016}, Creative Commons Attribution 4.0 International License. (e), (f) Reproduced with permission from Ref. \cite{Miyoshi2014}, article copyrighted by JPS \copyright 2014, The Physical Society of Japan.}
	\label{fig:10}
\end{figure}

Although FeSe does not order magnetically at ambient pressure \cite{Medvedev2009,McQueen2009,Bendele2010}, magnetic order is induced by pressures exceeding $\sim0.8$ GPa. The detailed determination of the pressure-temperature phase diagram is the result of several years of effort by a number of groups (Figs. \ref{fig:9}, \ref{fig:10}, \ref{fig:13}).  
Surprisingly, the structural transition at \mbox{$T_\mathrm{s}=90$ K} at ambient pressure decreases under pressure\cite{Miyoshi2014}, whereas the magnetic transition temperature $T_\mathrm{N}$ increases\cite{Bendele2010,Bendele2012,Terashima2015,Sun_2016}. These opposing trends are highly unusual for iron-based systems. {\red The resistance data as shown in Fig. \ref{fig:10} allows to follow this evolution rather well. The structural transition results in a small kink in the resistance, whereas the signature of the magnetic transition is an upward jump at low pressures and a downward jump at higher pressures.} Ultimately, the magnetic order has a dome-like pressure dependence with a maximum of $\sim45$ K around 4 GPa\cite{Sun_2016} (Fig. \ref{fig:9}(e)). 

High-resolution x-ray diffraction under pressure combined with time-domain M\"ossbauer spectroscopy showed that structural and magnetic phase transition lines actually merge around 1.6 GPa into a joint magneto-structural transition at higher pressures \cite{Kothapalli2016} (Fig. \ref{fig:9}(f)). NMR similarly demonstrates the decrease of $T_\mathrm{s}$ under pressure until structural and magnetic phase transitions merge into a combined first-order transition around 2 GPa\cite{Wang2016}. An important question is whether the magnetically ordered phase in FeSe is analogous to the stripe-type phase in other iron-based superconductors, as suggested by the merging of structural and magnetic transitions. This question will be discussed in section \ref{sec:pressure}.

The evolution of the superconducting transition in FeSe under hydrostatic pressure appears to be closely related to the pressure dependence of nematic and magnetic orders.
$T_\mathrm{c}$ of FeSe under pressure exhibits a complex three-step-like increase (Fig. \ref{fig:9}(c),(e)), as revealed by transport as well as by magnetization measurements \cite{Miyoshi2014,Sun_2016}. $T_\mathrm{c}$ initially increases under pressure and reaches a local maximum with $T_\mathrm{c}\approx 13$ K at 0.8 GPa, the pressure at which magnetic order sets in. Subsequently, $T_\mathrm{c}(p)$ shows a local minimum around 1.2 GPa and a plateau at $\sim2-5$ GPa. $T_\mathrm{c}$ reaches its maximum of 38.3 K close to 6 GPa, when magnetic order is suppressed \cite{Sun_2016}. Further application of pressure reduces $T_\mathrm{c}$. At pressures $\gtrsim 7-10$ GPa, tetragonal FeSe irreversibly transforms into a new crystallographic structure, identified as orthorhombic\cite{Kumar2010,Svitlyk2016,Margadonna2009} (previously sometimes as hexagonal\cite{Braithwaite2009,Medvedev2009}) with significantly reduced unit cell volume. This transformation likely leads to a loss of superconductivity\cite{Sun_2016}.

\subsection{Magnetic order under pressure}
\label{sec:pressure}
\begin{figure}
	\includegraphics[width=\textwidth]{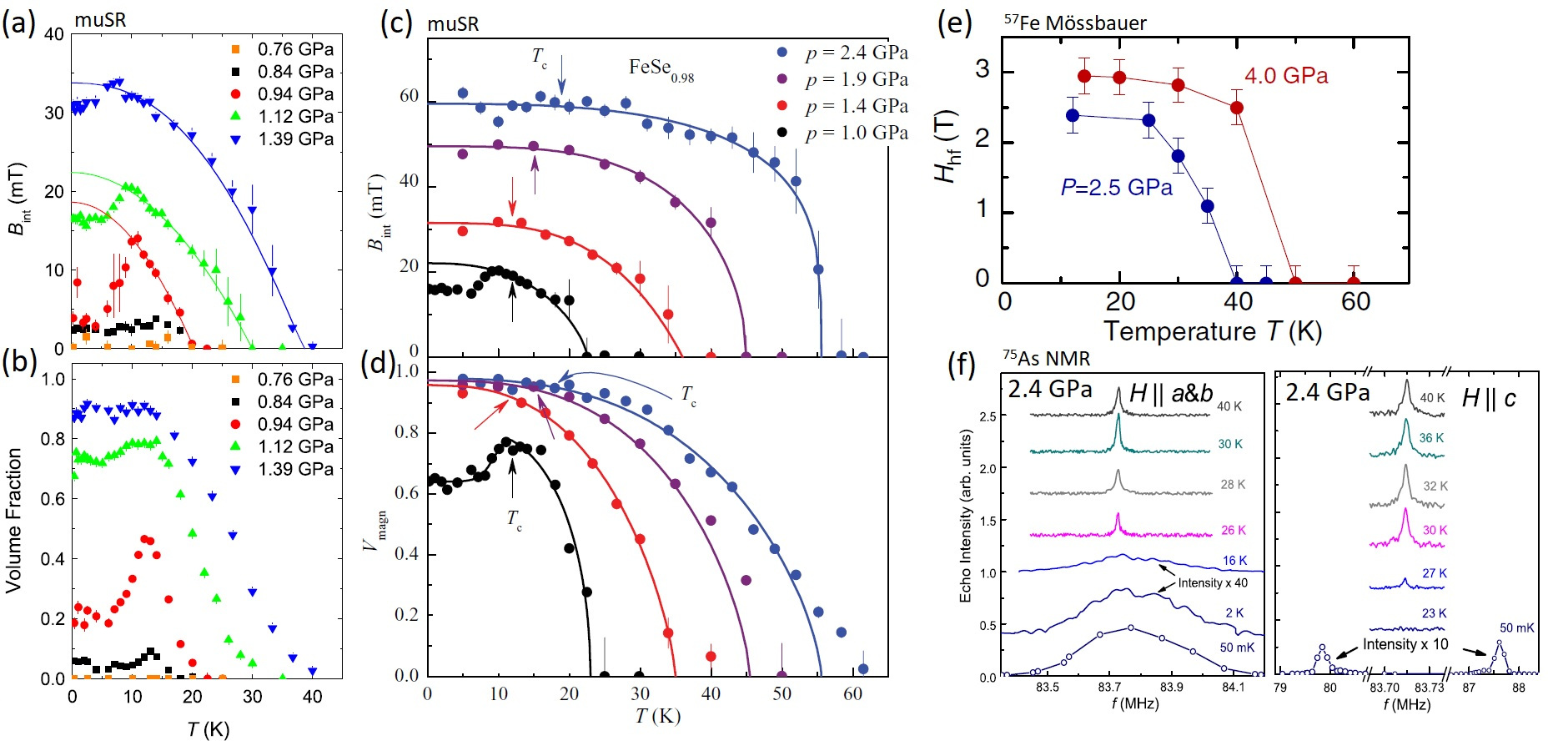}
	\caption{Magnetic order in FeSe under pressure. (a),(b) Temperature dependence of the internal magnetic field at the muon site and the magnetic volume fraction vs. temperature, respectively, at various pressures up to 1.39 GPa. (c),(d) Internal magnetic field at the muon site and magnetic volume fraction at pressures up to 2.4 GPa, respectively. (e) Temperature dependence of the magnetic hyperfine field at the iron-site at 2.5 GPa and 4.0 GPa. (f)$^{77}$Se NMR spectra of single-crystalline FeSe at 2.4 GPa under in-plane and $c$-axis magnetic fields. A hyperfine field at the As-site of 0.48 T oriented along the $c$-axis is revealed. (a), (b) Reproduced with permission from Ref. \cite{Bendele2010}, copyright 2010 American Physical Society. (c), (d) Reproduced with permission from Ref. \cite{Bendele2012}, copyright 2012 American Physical Society. (e) Adapted from Ref. \cite{Kothapalli2016}, Creative Commons Attribution 4.0 International License. (f) Reproduced with permission from Ref. \cite{Wang2016}, copyright 2016 American Physical Society.}
	\label{fig:FeSe_pressure_microscopic}
    \label{fig:13}
\end{figure}

{\red Magnetotransport\cite{Terashima2016II} and quantum oscillation\cite{Terashima2016} experiments under pressure demonstrate a significant reconstruction of the Fermi surface in the magnetic state, as expected for the onset of antiferromagnetic order.}
The most direct microscopic probe for the nature of magnetic order would be neutron diffraction. However, neutron experiments on FeSe under pressure have so far not succeeded to resolve magnetic signals. This is likely due to the small ordered magnetic moment. From the early muSR data (Fig. \ref{fig:13}(a)-(d)) an ordered moment of only $\sim0.2 \mu_B$\cite{Bendele2010,Bendele2012} is inferred at 2.5 GPa. Similarly, time-domain M\"ossbauer spectroscopy\cite{Kothapalli2016} is consistent with a small value of the ordered moment of the order of $\sim0.2$ $\mu_B$/Fe at $p=4$ GPa (Fig. \ref{fig:13}(e)).

High-resolution x-ray diffraction \cite{Kothapalli2016} demonstrates that the tetragonal symmetry is broken in the magnetic state, as is the case for stripe-type magnetic order in other iron-based systems. Moreover, an abrupt increase of the orthorhombic distortion in decreasing temperature through $T_\mathrm{N}$ is observed, reminiscent of the behavior of Co- and Rh-doped BaFe$_2$As$_2$ \cite{Kim2011} and a sign of cooperative coupling of the two types of order. $^{77}$Se NMR study of single crystalline FeSe under pressure\cite{Wang2016} likewise suggests stripe-type antiferromagnetic order. In particular, the observed $c$-axis oriented magnetic hyperfine field at the Se site (Fig. \ref{fig:13}(f),(g)) is consistent with stripe-type ordering and Fe-moments pointing along the $a$-axis\cite{Wang2016}, as in the 122-type iron-based materials. In addition, the in-plane/out-of-plane anisotropy of the spin-lattice relaxation rate of 1.5 has been taken as an indication of stripe-type magnetic fluctuations above $T_\mathrm{N}$ \cite{Wang2016}. However, observed changes of spectral weight might indicate a broad distribution of magnetic moments, or phase inhomogeneity in the magnetic state\cite{Wang2016}.
A muSR study on a crystal 'with preferred orientation'  under pressure \cite{Khasanov2016} is also consistent with stripe-type antiferromagnetic order.

An interaction between magnetic order and superconductivity is observed only in the low-pressure range. Upon decreasing the temperature below $T_c$, a decrease of the magnetic hyperfine field and volume fraction are observed in muSR on polycrystalline samples for $p \leq 1.4$ GPa \cite{Bendele2010,Bendele2012} (Fig. \ref{fig:FeSe_pressure_microscopic}(a)-(d)). 

\begin{figure}
	\includegraphics[width=\textwidth]{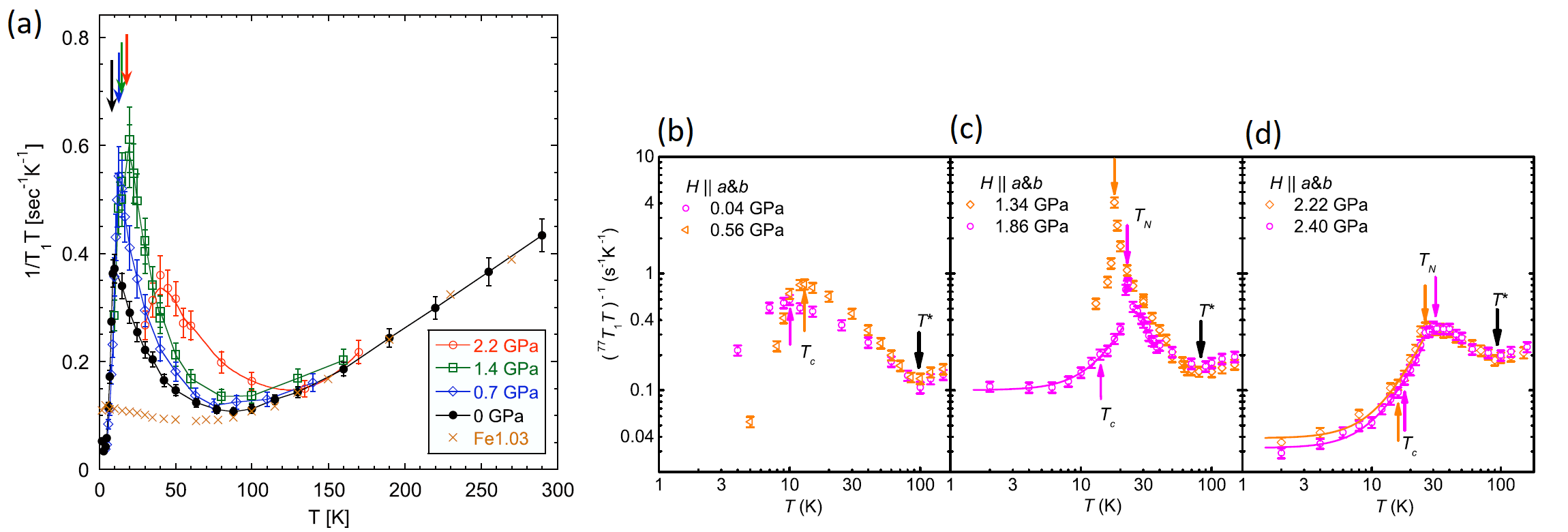}
	\caption{The $^{77}$Se spin-lattice relaxation rate under pressure. (a)  $^{77}$Se spin-lattice relaxation rate divided by temperature, $1/T_1T$, of polycrystalline FeSe under pressure. The upturn below $T^*\sim100$ K is associated with the emergence of low-energy spin fluctuations. (b)-(d) $1/T_1T$ at various pressures, indicating a first-order magnetic transition at high pressures. The onset of enhanced relaxation rate, $T^*$, is nearly pressure independent. (a) Reproduced with permission from Ref. \cite{Imai2009}, copyright 2009 American Physical Society. (b)-(d) Reproduced with permission from Ref. \cite{Wang2016}, copyright 2016 American Physical Society.}
    \label{fig:14}
\end{figure}

The temperature evolution of the spin-lattice relaxation rate {\red divided by temperature} $1/T_1T$ in NMR under pressure (Fig. \ref{fig:14}) allows to study magnetic fluctuations and the evolution of the character of the magnetic transition under pressure as well. Measurable magnetic fluctuations onset below a temperature $T^*$ that is essentially pressure independent\cite{Wang2017,Wiecki2017}, strongly suggesting that the coincidence of $T^*$ and $T_\mathrm{s}$ at ambient pressure is accidental. {\red Interestingly, the onset of these low-energy spin fluctuations correlates with the onset of local static nematicity, suggesting their cooperative interplay\cite{Wiecki2017}. This extended region of static inhomogeneous nematicity in the temperature-pressure phase diagram of FeSe was also reported in Ref. \cite{Wang2017III}.} On decreasing temperature at intermediate pressures, $1/T_1T$ shows a clear divergence, indicative of a second-order transition at $T_\mathrm{N}$ when $T_\mathrm{s}>T_\mathrm{N}$. However, this divergence is absent at higher pressures, indicating a first-order magneto-structural transition at $T_\mathrm{s}=T_\mathrm{N}$\cite{Wang2016}.

\section{Superconductivity\label{sec:superconductivity}}
\begin{figure}
	\includegraphics[width=\textwidth]{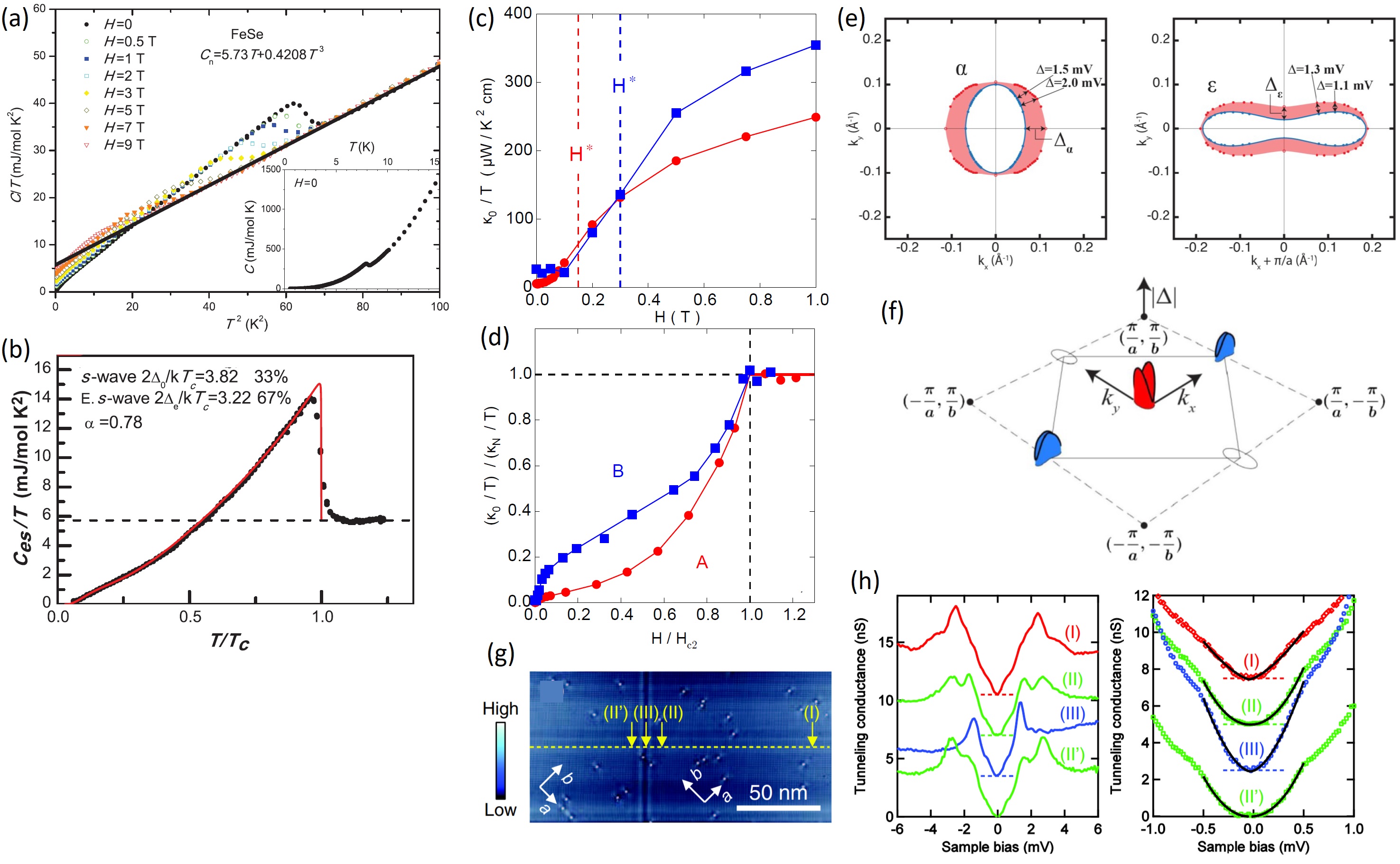}
	\caption{Experimental results concerning the superconducting gap structure of FeSe. (a) Specific heat of single-crystalline FeSe. (b) Electronic specific heat of FeSe and fitted BCS-type model consisting of a sum of an isotropic s-wave gap and a nodeless anisotropic (extended s-wave) gap. (c), (d) Thermal conductivity of two samples of FeSe indicating the absence of nodes. (e) Gap structure of the central hole and elongated electron pockets of FeSe as determined by Bogoliubov quasiparticle interference. A pronounced two-fold anisotropy of the gap is observed for the electron pocket. (f) Illustration of this gap structure; the color-code indicates the sign change between the Fermi surface pockets \cite{Sprau2017}. (g) STM image of a domain boundary of FeSe. (h) Tunneling spectra taken a several points across this domain boundary as indicated in (g). The low-energy part of the tunneling spectra suggests possible changes of the gap structure as a function of position. (a), (b) Reproduced with permission from Ref. \cite{Lin2011}, copyright 2011 American Physical Society. (c), (d) Reproduced with permission from Ref.  \cite{Bourgeois-Hope2016}, copyright 2016 American Physical Society. (e), (f) From Ref. \cite{Sprau2017}. Reprinted with permission from AAAS. (g), (h) Reproduced from Ref. \cite{Watashige2015}, Creative Commons Attribution 3.0 License.}
	\label{fig:20}
\end{figure}

Superconductivity in FeSe has many intriguing and complex features. 
The superconducting transition temperature of bulk FeSe at ambient pressure is about \mbox{$8-9$ K}, which is modest for iron-based superconductors. However, there are numerous ways to increase this $T_\mathrm{c}$ significantly. Application of hydrostatic pressure or intercalation of ions or molecules increases $T_\mathrm{c}$ to 35-45 K\cite{Mizuguchi2008,Medvedev2009,Margadonna2009,Ying2011II,Hrovat2015,Rebec2017}. In monolayer thin films of FeSe on SrTiO$_3$, $T_\mathrm{c}$ can reach 60-80 K\cite{Wang2012}, and by some accounts even more than 100 K\cite{Ge2015}, as recently reviewed in Refs. \cite{Sadovskii2016,Wang2017,Huang2017}.

The enormous tunability of $T_\mathrm{c}$ of FeSe hints at an unconventional origin of superconductivity in the compound.  The observation of a spin-resonance mode (Fig. \ref{fig:8}(e),(f)) is consistent with a spin-fluctuation-mediated
sign-changing pairing mechanism \cite{Wang2015}. Furthermore, the analysis of STM data\cite{Sprau2017} with a phase sensitive method\cite{Hirschfeld15,Martiny2017} found evidence for a sign change of superconducting gap between electron and hole Fermi surfaces, the presence of in-gap resonances at impurities was interpreted with the same conclusion\cite{Jiao2017}. Recently, a strong electron-phonon interaction in FeSe was deduced from a pump-probe experiment and the importance of a cooperative interplay between electron-electron and electron-phonon interactions for superconductivity in FeSe was suggested\cite{Gerber17}.

The extremely small Fermi-surfaces of FeSe together with the 'typical' values for the superconducting gaps of 2-3 meV mean that the ratio of $\Delta$ to $E_{\mathrm{F}}$ in FeSe is as high as $0.1-1$. This places FeSe into the regime of a (multiband) BCS-BEC crossover, which has been intensively discussed\cite{Kasahara2014,Kasahara2016,Watashige_2017}. {\red In particular, strong superconducting fluctuations associated with preformed Cooper pairs have been reported \cite{Kasahara2016}}. Fe(Se,Te) with $\sim40\%-60\%$ Te content appears to be in a similar regime \cite{Lubashevsky2012,Okazaki2014}. 

The superconducting gap structure of FeSe has been subject of intense study; some recent results are summarized in Fig. \ref{fig:20}. Signatures of multigap superconductivity are generally observed \cite{Lin2011,Abdel-Hafiez2013,Bourgeois-Hope2016,Naidyuk2017,Sun_2017arXiv}.
The dimensionless specific heat jump at $T_\mathrm{c}$ of $\Delta C/\gamma_nT_\mathrm{c} = 1.65$ indicates moderately or strong coupling superconductivity \cite{Lin2011}. Early specific heat data in single crystals are consistent with the absence of nodes in the superconducting gap\cite{Lin2011}. However, from STM data on thin films\cite{Song2012,Song2012II} and on single crystals\cite{Kasahara2014}, it was suggested that clean FeSe is a nodal superconductor. Other more recent studies including STM\cite{Sprau2017,Jiao_2017}, specific heat\cite{Jiao_2017}, thermal conductivity\cite{Bourgeois-Hope2016,Watashige_2017} and the London penetration depth\cite{Li2016,Teknowijoyo2016} indicate, however, that FeSe is fully gapped superconductor, albeit with deep gap minima. The specific-heat data in Refs. \cite{Wang2017II} and \cite{Chen2017} are consistent with nodal superconductivity or deep gap minima. A recent detailed study of the field-angle dependent specific heat of FeSe\cite{Sun_2017arXiv} proposes three distinct superconducting gaps, of which the two smallest ones are anisotropic, and the smallest possibly nodal. 

It is likely that small modifications in sample preparation, leading to a change of the precise Fe:Se ratio, or to internal stresses can induce variations of the gap structure from nodal to nodeless with deep minima of the gaps. It was shown that changes in sample preparation can have a significant impact on $T_\mathrm{c}$, which can vary between \mbox{$3-9$ K} for samples prepared with slightly different starting compositions and temperature profiles \cite{Boehmer2016}. Effects of impurity scattering can explain the variation of $T_\mathrm{c}$ and at the same time the filling in of low-energy density of states. {\red Indeed, a proton irradiation study concludes that if there are gap nodes in FeSe, these are "symmetry-unprotected"\cite{Sun2017II}}. Furthermore, the superconducting gap appears to also become ``more'' nodeless close to an orthorhombic domain boundary\cite{Watashige2015}. Thus, it might be possible that nodal and nodeless spatial regions coexist in a bulk sample.

Very interesting is the pronounced two-fold anisotropy of the superconducting properties in FeSe. A sizable in-plane anisotropy of the superconducting coherence lengths, as shown by the spatial extent of the vortex cores, was observed early in thin films\cite{Song2011} and confirmed by measurements on bulk single crystals \cite{Watashige2015}. A clear two-fold anisotropy of the superconducting gaps was demonstrated by high-resolution Bogoliubov quasiparticle interference \cite{Sprau2017}. ARPES on lightly S-doped FeSe likewise revealed a significant 2-fold anisotropy of the superconducting gap\cite{Xu2016}. 

The pronounced two-fold anisotropy of the superconducting gap structure, given that the crystal structure is distorted only very slightly from tetragonal, suggests a strong link between nematicity and superconductivity in FeSe. In the prototypical 122-type superconductors, the interaction between superconductivity and stripe-type magnetic order leads to a sizeable decrease of orthorhombic distortion and ordered magnetic moment below $T_\mathrm{c}$\cite{Nandi2010,Fernandes2010II}. In contrast, the relation between orthorhombic lattice distortion and superconductivity in FeSe is very subtle. The small changes of the lattice parameters at $T_\mathrm{c}$ were studied using high-resolution dilatometry in Ref. \cite{Boehmer2013}. It was found that the $a$ and $b$ axes change in a very similar manner at the superconducting transition, so that the effect of $T_\mathrm{c}$ on the orthorhombic distortion is only a slight curvature change\cite{Boehmer2013}. As can be inferred via a thermodynamic relation, this indicates that $T_\mathrm{c}$ is unaffected by small changes of $\delta$. Interestingly, in lightly S-doped FeSe, the same method indicates that superconductivity couples \emph{cooperatively} with orthorhombic distortion\cite{Wang2017II}. This means that an increase of $\delta$ would result in an increase of $T_\mathrm{c}$, a behavior not seen in any other iron-based superconductor so far.

In contrast to the structural distortion, the in-plane anisotropy of the NMR spectra at ambient pressure actually decreases slightly below $T_\mathrm{c}$ \cite{Baek2015,Wiecki2017}. This was proposed to be a sign of competition between electronic nematicity and superconductivity \cite{Baek2015}. Interestingly, no such competition is visible in the lattice parameters\cite{Boehmer2013}.

Interestingly, $T_\mathrm{c}$ appears to be generally suppressed by presence of the pressure-induced magnetic order in FeSe. In particular, $T_\mathrm{c}(p)$ exhibits a local maximum at $p\sim0.8$ GPa (the point of emergence of magnetic order) and a global maximum at $p\sim6$ GPa, when $T_\mathrm{N}$ is reduced to below $T_\mathrm{c}$. However, in the pressure range \mbox{$\sim2-6$ GPa}, both $T_\mathrm{c}$ and $T_\mathrm{N}$ appear to increase under pressure, which is unexpected if magnetic order and superconductivity compete strongly. A decrease of magnetic hyperfine field below the superconducting transition has been demonstrated only in the low pressure range $p\sim0.8-1.4$ GPa\cite{Bendele2010,Bendele2012}. Furthermore, no change of orthorhombic distortion at $T_\mathrm{c}$ is resolved at any pressure \cite{Kothapalli2016}. These observations may indicate a much weaker interaction between superconductivity and magnetic order than in other iron-based systems. {\red Experimental studies of the superconducting gap structure under pressure are exceedingly difficult and have not been reported to date}.

There is a possibility that superconductivity in FeSe does not even coexist microscopically with magnetic order above a certain pressure value, which would explain why no signature of $T_\mathrm{c}$ is observed in the lattice parameters and hyperfine fields. The transport and magnetic signatures of superconductivity under pressure may arise from only a small volume fraction of the sample. This was suggested based on NMR data which do not find any change of the spin-lattice relaxation rate $1/T_1$ below $T_\mathrm{c}$ under pressure \cite{Wang2016}. Note that in 122-type materials with a strong first-order magneto-structural transition like Ca(Fe$_{1-x}$Co$_x$)$_2$As$_2$, microscopic coexistence between magnetic order and superconductivity is not observed either \cite{Ran2012}. Recently, the absence of bulk superconductivity in the magnetically ordered part of the phase diagram has indeed been indicated by AC magnetic susceptibility in slightly sulfur-substituted FeSe\cite{Yip2017}. Future experimental studies are needed to reveal to which extent superconductivity can coexist with magnetic order in FeSe under pressure.

\section{Effects of chemical substitutions in FeSe}
\label{sec:substitution}
\begin{figure}
	\includegraphics[width=\textwidth]{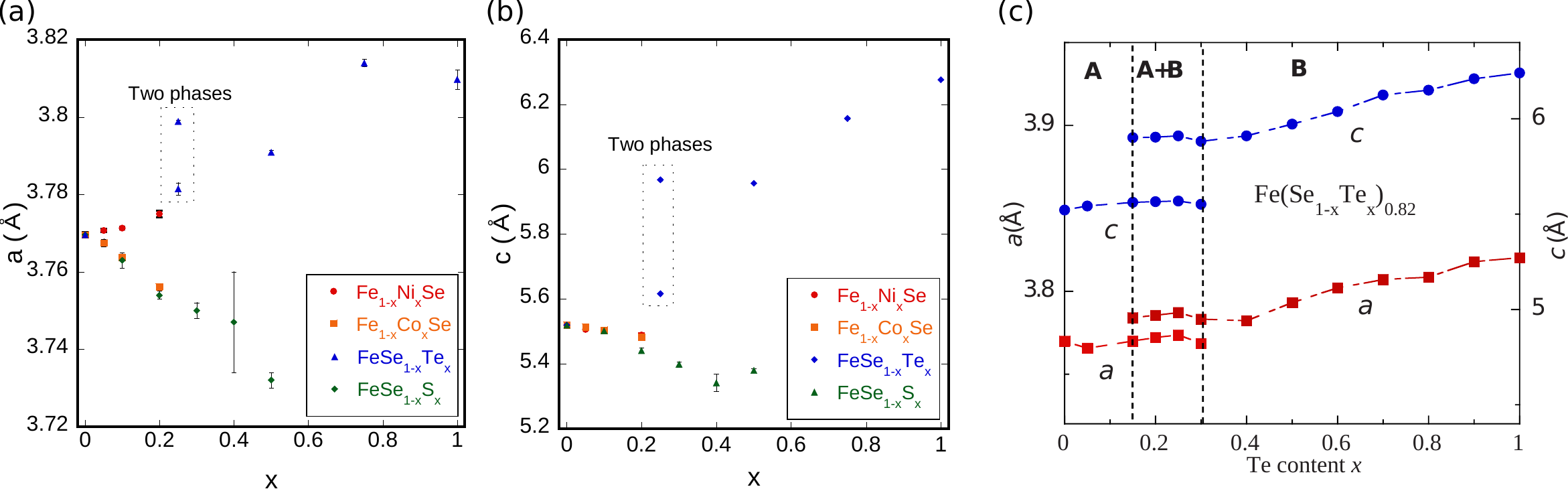}
	\caption{Evolution of the lattice parameters of substituted FeSe. (a), (b) $a$- and $c$-lattice parameter, respectively, of polycrystalline FeSe as a function of Ni, Co, Te, and S. A two-phase region in the Fe(Se,Te) series is indicated. (c) Evolution of lattice parameters in the Fe(Se,Te) series showing the two-phase region between $\sim 10-30\%$ Te content. (a), (b) Reproduced with permission from Ref. \cite{Mizuguchi2009}, articles copyrighted by JPS \copyright 2009 The Physical Society of Japan. (c) Reproduced with permission from Ref. \cite{Fang2008II}, copyright 2008 American Physical Society.}
	\label{fig:substitutions}
\end{figure}

\begin{figure}
	\includegraphics[width=\textwidth]{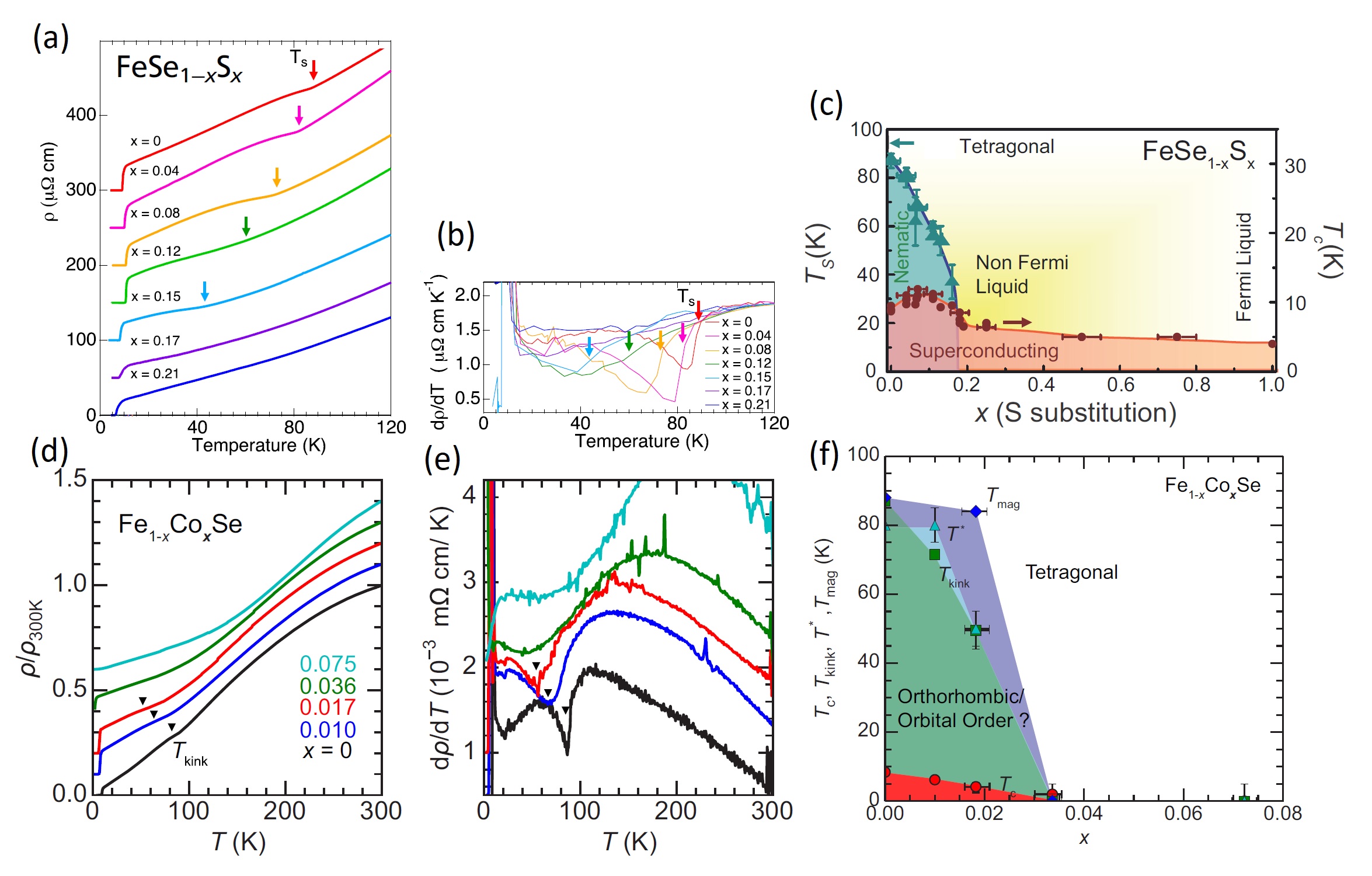}
	\caption{Electrical resistance and phase diagrams of Co- and S-substituted FeSe single crystals. (a), (b) Resistance and temperature derivative of the resistance of S-substituted FeSe, respectively. Data in (a) are offset vertically. (c) Phase diagram of FeSe$_{1-x}$S$_x$. Note that $T_s$ and $T_c$ are plotted on the left and right temperature axis, respectively. (d), (e) Resistance and temperature derivative of the resistance of Co-substituted FeSe, respectively. Data in are offset vertically. (f) Phase diagrams of Co-substituted FeSe. (a), (b) Reproduced from Ref. \cite{Hosoi2016}. (c) Reproduced with permission from Ref. \cite{Reiss2017}, copyright 2017 American Physical Society. (d)-(f) Reproduced with permission from Ref. \cite{UrataKontaniTanigaki2016PRB_FeSe-Co-doping-pairing}, copyright 2016 American Physical Society.}
	\label{fig:17}
\end{figure}

Chemical substitutions are a common way of tuning iron-based materials\cite{Canfield2010}. However, the variety of successful chemical substitution in FeSe is rather limited as compared to many other iron-based systems. 
Initially, substitution of Te and S for Se, and of Co and Ni for Fe have been studied in some detail using polycrystalline samples and the Fe(Se,Te) system has received particular attention\cite{Mizuguchi2009}, see Fig. \ref{fig:substitutions} (a), (b). A tetragonal-to-orthorhombic structural transition appears at a reduced \mbox{$T_\mathrm{s}=40$ K} in Fe$_{1.03}$Se$_{0.57}$Te$_{0.43}$ \cite{Gresty2009}. The superconducting transition temperature reaches a maximum of \mbox{$T_\mathrm{c}=14$ K} at approximately 50\% Te content\cite{Mizuguchi2010II}. The FeTe endmember of the series orders antiferromagnetically, but with a double-stripe structure---different from most iron-based systems\cite{Bao2009}. This system is complex due to a significant amount of excess Fe leading to a rich Fe$_{1+\delta}$Te phase diagram\cite{Roessler2011,Koz2013}. Furthermore, the Se and Te ions occupy the same Wyckoff position in the lattice, but have different heights above the iron plane, entailing a significant amount of structural disorder \cite{Tegel2010}. A miscibility gap between 10\% and 30\% Te content has been demonstrated \cite{Fang2008II}, see Fig.\ref{fig:substitutions} (c) . Fe(Se,Te) has been reviewed previously\cite{Mizuguchi2010} and will not be discussed in further detail here. Attempts to grow Te-substituted single crystals of FeSe using vapor transport have failed\cite{Karlsson2015} and the miscibility gap at low Te content\cite{Fang2008II} makes connecting the Fe(Se,Te) series with pure FeSe difficult.  

For vapor-grown single crystals, the two main kinds of substitution are S for Se and Co for Fe. Sulfur substitution is considered to act as chemical pressure, since it leads to a decrease of the lattice parameters\cite{Matsuura2017} and sulfur is isovalent to Se, see Fig. \ref{fig:substitutions}. Cobalt substitution is more likely to introduce additional charge carriers. Nevertheless, both Co and S substitution suppress the structural transition at $T_\mathrm{s}$\cite{UrataKontaniTanigaki2016PRB_FeSe-Co-doping-pairing,Urata2016,Abdel-Hafiez2016,Hosoi2016,Watson2015III}. The critical concentrations, at which $T_\mathrm{s}$ is completely suppressed, are approximately 16\% S content and only 3\% Co content\cite{UrataKontaniTanigaki2016PRB_FeSe-Co-doping-pairing}, see Fig. \ref{fig:17}. $T_\mathrm{c}$ initially increases with S substitution and reaches a broad maximum of $11-12$ K at $\approx8\%$ S content\cite{Hosoi2016,Watson2015III,Coldea2016,Reiss2017}, well inside the orthorhombic phase. Further S substitution decreases $T_\mathrm{c}$ moderately to $\approx6$ K at 20\% S content. Co substitution, conversely, decreases $T_\mathrm{c}$ monotonically and superconductivity is completely suppressed concomitantly with $T_\mathrm{s}$ at about 3\% Co content\cite{UrataKontaniTanigaki2016PRB_FeSe-Co-doping-pairing}. No sign of magnetic order is observed at ambient pressure for either kind of substitution. Early NMR work has demonstrated the suppression of the low-temperature magnetic fluctuations at 10\% Co substitution\cite{Kotegawa2012}. 

Since the orthorhombic phase of FeSe is a rare example of a nematic phase without magnetic order in the iron-based systems, suppressing $T_\mathrm{s}$ allows for the study of a pure nematic quantum critical point (QCP) \cite{Hosoi2016}. Indeed, a Fermi-liquid-like $T^2$ behavior of resistivity is observed only at high sulfur concentrations, whereas the resistivity at intermediate sulfur concentrations suggests non-Fermi-liquid behavior close to the endpoint of the nematic phase\cite{Reiss2017}, see Fig. \ref{fig:17}. This strongly resembles the behavior of the resistivity in BaFe$_2$(As$_{1-x}$P$_{x}$)$_2$\cite{Kasahara2010}, which has been discussed in terms of a quantum criticality \cite{Kasahara2010,Nakai2010II,Hashimoto2013,Putzke2014,Hayes2016}. However, in both S and Co substituted FeSe, $T_\mathrm{c}$ is not maximum at the QCP, in contrast to BaFe$_2$(As$_{1-x}$P$_{x}$)$_2$. 

The nematic susceptibility, as probed by the strain-dependence of the resistivity, was studied across the possible nematic quantum critical point in the Fe(Se,S) system\cite{Hosoi2016}. The nematic susceptibility was found to diverge with an approximate Curie-Weiss law for all compositions and the Weiss temperature changes sign at the critical concentration of $\approx15\%$ S content. Notably, the amplitude of elastoresistivity is strongly enhanced close to this critical point. This behavior is strongly reminiscent of the behavior of Ba(Fe$_{1-x}$Co$_x$)$_2$As$_2$ and has been discussed as a signature of a nematic quantum critical point\cite{Chu2012}.  

The evolution of the electronic structure of FeSe with S substitution has been studied in detail using ARPES\cite{Watson2015III,Reiss2017}, quantum oscillations\cite{Coldea2016} and STM\cite{Hanaguri2017}. The distortion of the electron Fermi-surfaces was shown to be reduced by increasing S content as the orthorhombicity is gradually suppressed\cite{Watson2015III}. Furthermore, the Fermi surfaces generally become larger upon S substitution \cite{Coldea2016} and electronic correlations weaken\cite{Reiss2017}.
{\red The pronounced two-fold anisotropy of the superconducting gap of FeSe is clearly related to its orthorhombic crystal structure. Therefore, it is interesting to study the evolution of the superconducting gap in FeSe$_{1-x}$S$_x$, when the ground state evolves from orthorhombic to tetragonal. An analysis of specific heat and thermal conductivity found that even in tetragonal FeSe$_{1-x}$S$_x$ the superconducting gap is still highly anisotropic and possibly nodal\cite{Sato2017}. However, the recent STM study found evidence for two distinct superconducting pairing states in the orthorhombic and tetragonal samples, respectively\cite{Hanaguri2017}.}

Pure FeS with the same PbO-type tetragonal crystal structure as FeSe is a superconductor with $T_\mathrm{c}=5$ K\cite{Lai2015}. This crystallographic phase is, however, only stable below $200-250^\circ$C\cite{Borg2016}. FeS has not been grown out of the vapor phase, but either by hydrothermal synthesis from iron powder and sulfide solution\cite{Lai2015,Pachmayr2015} or by deintercalation of K-Fe-S\cite{Borg2016}. Likely due to this challenge, the complete series Fe(Se,S) could not yet be studied. The normal and superconducting properties of the multiband system FeS were studied in detail\cite{Lin2016,Yang2016,Ying2016,Xing2016}, indicating nodal superconductivity\cite{Ying2016,Xing2016}. 
The superconducting transition temperature is monotonically suppressed under pressure\cite{Holenstein2016}.
A recent comprehensive inelastic neutron diffraction, quantum oscillations and elastoresistivity study\cite{Man2017} suggested FeS to be a weakly correlated analogue to FeSe. 

\begin{figure}
	\includegraphics[width=\textwidth]{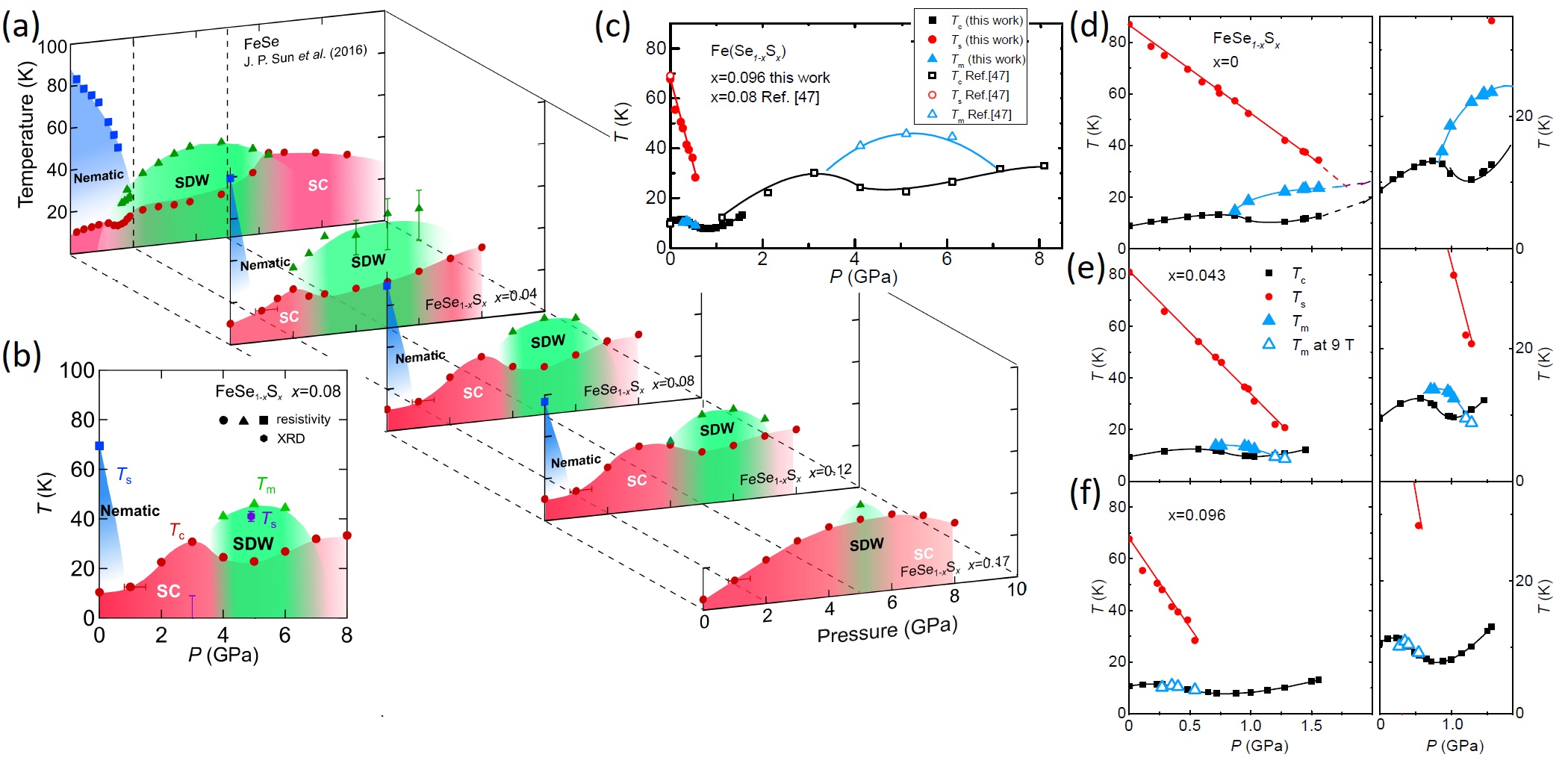}
	\caption{Phase diagram of FeSe with combined S substitution and applied pressure (a) Pressure-temperature phase diagrams of Fe(Se$_{1-x}$S$_x$) for $0 \leq x \leq 0.17$ showing the shrinking of the pressure-induced magnetic dome and its detaching from the nematic phase with increased sulfur content $x$. (b) View of the pressure-temperature phase diagram of FeSe$_{0.92}$S$_{0.08}$. $T_\mathrm{c}$ is maximized in the absence of magnetic and nematic order. (d)-(f) Temperature pressure phase diagrams of Fe(Se$_{1-x}$S$_x$) for small $0 \leq x \leq 0.096$ in the low-pressure range. The presence of a small dome of magnetic order at low pressures, that is rapidly suppressed for increasing $x$, is suggested. (c) Combination of data from Ref. \cite{Matsuura2017} (labeled ``Ref. [47]'') and \cite{Xiang2017} (labeled ``this work'') to a detailed $p-T$ phase diagram of Fe(Se$_{1-x}$S$_x$) for $x\approx0.09$. (a), (b) Reproduced from Ref. \cite{Matsuura2017}, Creative Commons Attribution 4.0 International License. (c)-(f) Reproduced with permission from Ref. \cite{Xiang2017}, copyright 2017 American Physical Society.}
	\label{fig:21}
\end{figure}

The combination of chemical pressure induced by sulfur substitution and physical pressure leads to particularly interesting phase diagrams (Fig. \ref{fig:21}), which may further elucidate the phase interplay in FeSe. The structural transition at $T_\mathrm{s}$ is gradually suppressed by both chemical and physical pressure. In contrast, the pressure range over which magnetic order occurs is reduced on increasing the sulfur content, but the maximum $T_\mathrm{N}$ changes only slightly. In consequence, for a certain range of sulfur content, nematic order and magnetic order occur in detached parts of the temperature-pressure phase diagram of Fe(Se,S) (Fig. \ref{fig:21} (a)-(c)). Furthermore, samples with low sulfur content $<10\%$ exhibit an additional small dome of likely magnetic order in the low pressure range (Fig. \ref{fig:21} (e),(f)) \cite{Xiang2017}. This phase remains below $T_\mathrm{s}$ on increasing sulfur content and its emergence under pressure coincides with the local maximum of $T_\mathrm{c}$ (that occurs at 0.8 GPa in pure FeSe) for all studied compositions of Fe(Se,S) \cite{Xiang2017}. The superconducting upper critical field, $H_{c2}$, displays an anomaly at this point\cite{Xiang2017}, similar to the anomaly of $H_{c2}$ at the emergence of magnetic order in pure FeSe\cite{Kaluarachchi2016,Kang2016II}. Notably, $T_\mathrm{c}$ is of the order of 30-35 K whenever superconductivity sets in within a tetragonal, paramagnetic phase \cite{Matsuura2017}. This observation supports the picture of competing superconductivity and magnetic order.

\section{Theoretical mechanisms\label{sec:theory}}
\begin{figure}[tb]
	\includegraphics[width=\textwidth]{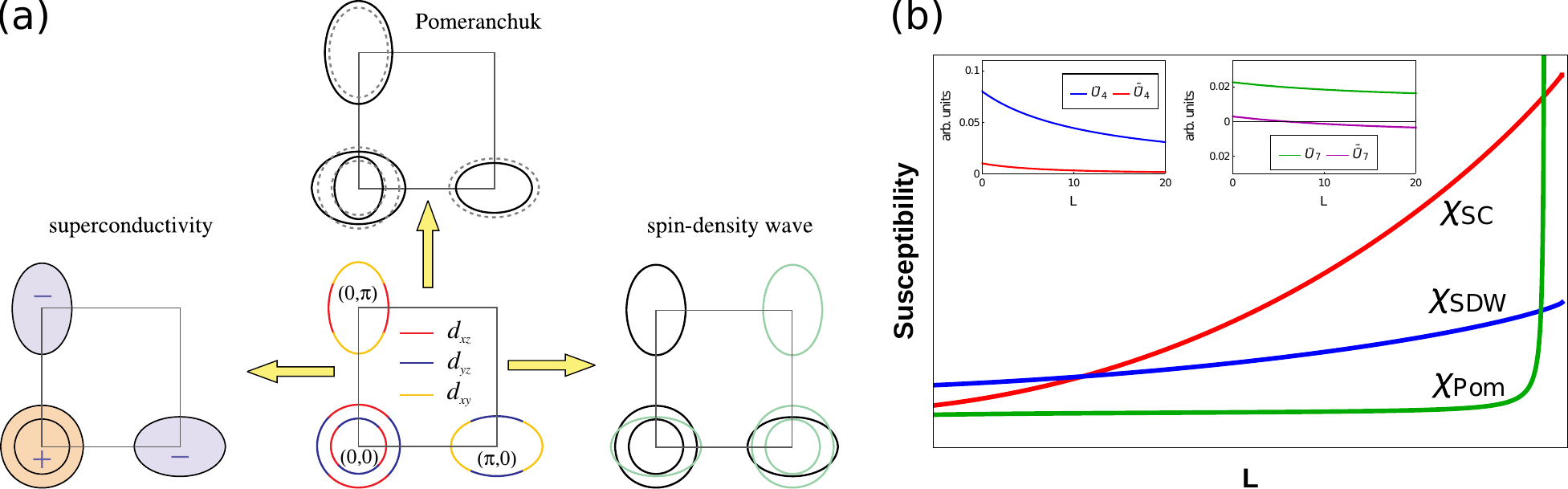}
	\caption{Potential instabilities at low energies as revealed by RG analysis. (a) The Fermi surface topology of Fe-based superconductors (orbital content plotted in color) allows for three electronic instabilities as indicated in the pictures on the left ($s_\pm$ superconductivity), right (stripe SDW magnetism) and top (nematicity, breaking of $C_4$ lattice rotational symmetry). (b) The flow of the susceptibilities of the 3 instabilities as function of the RG parameter $L$. While the susceptibilities in the superconducting channel and the Pomeranchuk channel diverge, thus leading to such an order, the one in the SDW channel increases, but remains finite as $L$ approaches the scale $L_0$ where the flow has to stop (slightly outside the plot range). Insets: Representative RG flow of some of 10 decoupled interactions as worked out in Ref. \cite{Xing17}; some of them flow to zero ($\tilde U_4$ and $\tilde{\tilde U_4}$), others flow to small but finite values ($\tilde U_7$ and $\tilde{\tilde U_7}$). (a) Reproduced from Ref. \cite{Chubukov2016}, Creative Commons Attribution 3.0 License. (b) Reproduced with permission from Ref. \cite{Xing17}, copyright 2017 American Physical Society.}
	\label{fig:theory0}
\end{figure}

Immediately after the discovery of the iron-based superconductors, ab-initio methods were applied to
calculate the basic properties of their electronic structure\cite{Eschrig09,Miyake_cRPA,Subedi2008}.
The general topology of the expected Fermi surface sheets and the nature of the low-energy electronic states as Fe-d states are in agreement with experimental observations\cite{Paglione2010,Maletz2014,Watson2015}. However, discrepancies between the results from DFT investigations and experimental results are particularly pronounced in the case of FeSe: In contrast to the theoretical results, the experimental Fermi surface sheets in FeSe are extremely small\cite{Maletz2014}. Furthermore, obtaining a stable orthorhombic crystal structure with DFT approaches requires imposing finite magnetic order\cite{Glasbrenner2015} which is (at zero pressure) experimentally absent.

The electronic structure of the iron-based superconductors was found to be only moderately correlated; much less than, for example, in the cuprates. Therefore, DMFT methods are a suitable approach to calculate electronic properties of iron-based superconductors. These methods have been applied to FeSe early on\cite{Aichhorn2010}. The general trend of band shifts at the $\Gamma$ and the M-point (red/blue shift) relative to the DFT result was obtained. It makes the Fermi surface sheets smaller, thus closer to the experimentally observed sizes\cite{Yin2011}. For a review on ab-initio perspectives see Ref. \cite{Guterding2017}. For FeSe in particular, such a DMFT based approach can account for the correct crystallographic $z_{\mathrm{Se}}$ position when relaxing the structure\cite{Haule2016}, and finds the so-called
lower Hubbard-bands which have been observed in ARPES measurements \cite{Evtushinsky2016,Watson2017}.

Features of the magnetic fluctuations have been calculated as well\cite{Yin2014}, but the nematic order and the resulting orthorhombic crystal structure are difficult to obtain from ab-initio based methods. This is also true for the extremely small Fermi surface sheets in FeSe\cite{Maletz2014,Watson2017}. Thus, various aspects of the material in the normal state and the superconducting state have been examined by model-based calculations where the specific choice of the parameters
needs to be motivated.

\begin{figure}[tb]
	\includegraphics[width=\textwidth]{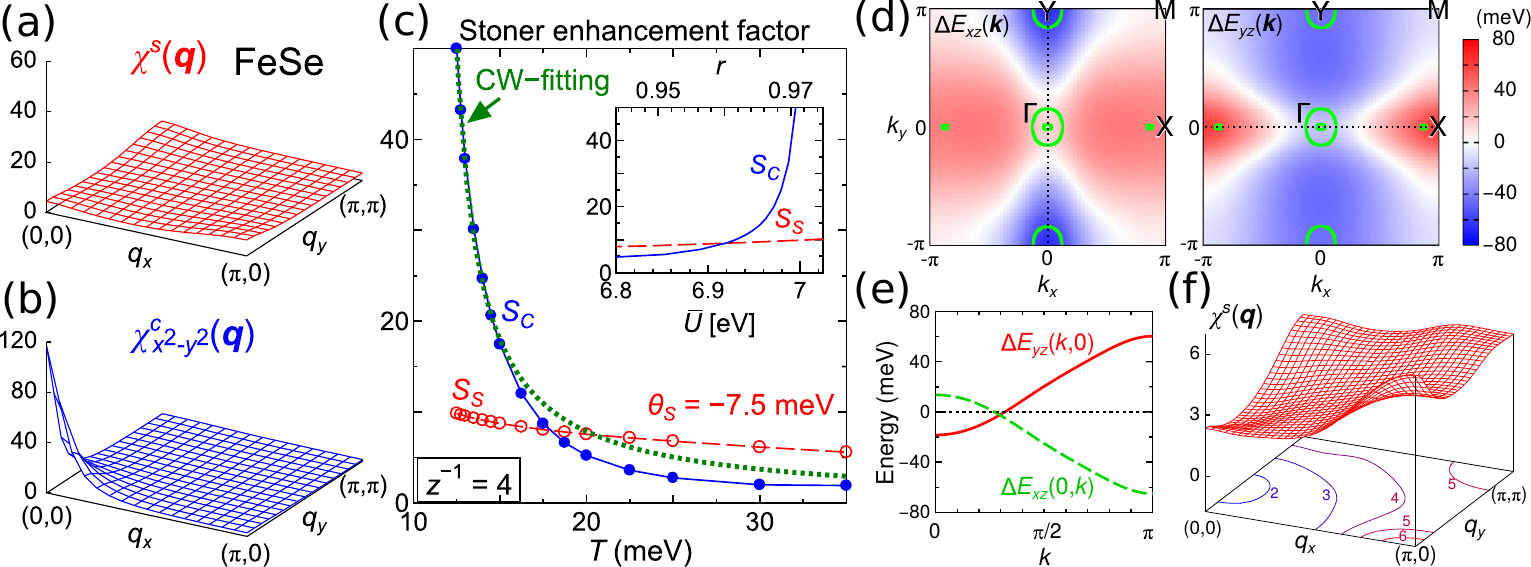}
	\caption{Nematic order in FeSe as found from a strong-coupling effect when working out the self-consistent vertex correction approach. (a) Spin susceptibility $\chi^s(q)$ as calculated with quasiparticle mass renormalization $z=1$ and interaction strength $r=0.25$. (b) Charge susceptibility $\chi^c(q)$ for the same parameters as in (a) exhibiting a large peak at $q=(0,0)$. (c) Stoner enhancement factors for the ferro-orbital order $S_C$ and the antiferromagnetic order $S_S$ which drive the respective order once becoming large. Setting the quasiparticle mass renormalization $z^{-1}=4$, $S_C$ becomes diverging. Inset: Enhancement factors as function of $r$ showing a divergence close to $r=0.97$. (d) Self-consistent solution of the orbital polarization, i.e. the orbital shifts $\Delta E_{xz}(k)$ and $\Delta E_{yz}(k)$ plotted in the Brillouin zone. The model exhibits an orbital order at $T=50\,\mathrm{meV}$ yielding a $C_2$ symmetric shape of the Fermi surface. (e) The orbital polarization $\Delta E_{yz}$ ($\Delta E_{xz}$) plotted along the $k_x$ ($k_y$) axis making the sign change of the orbital order evident. (f) The spin susceptibility $\chi^s(q)$ becomes $C_2$ symmetric in the orbital ordered state. Adapted from Ref.\cite{Yamakawa_PRX_16}, Creative Commons Attribution 3.0 License.}
	\label{fig:theory1}
\end{figure}

\begin{figure}[tb]
	\includegraphics[width=\textwidth]{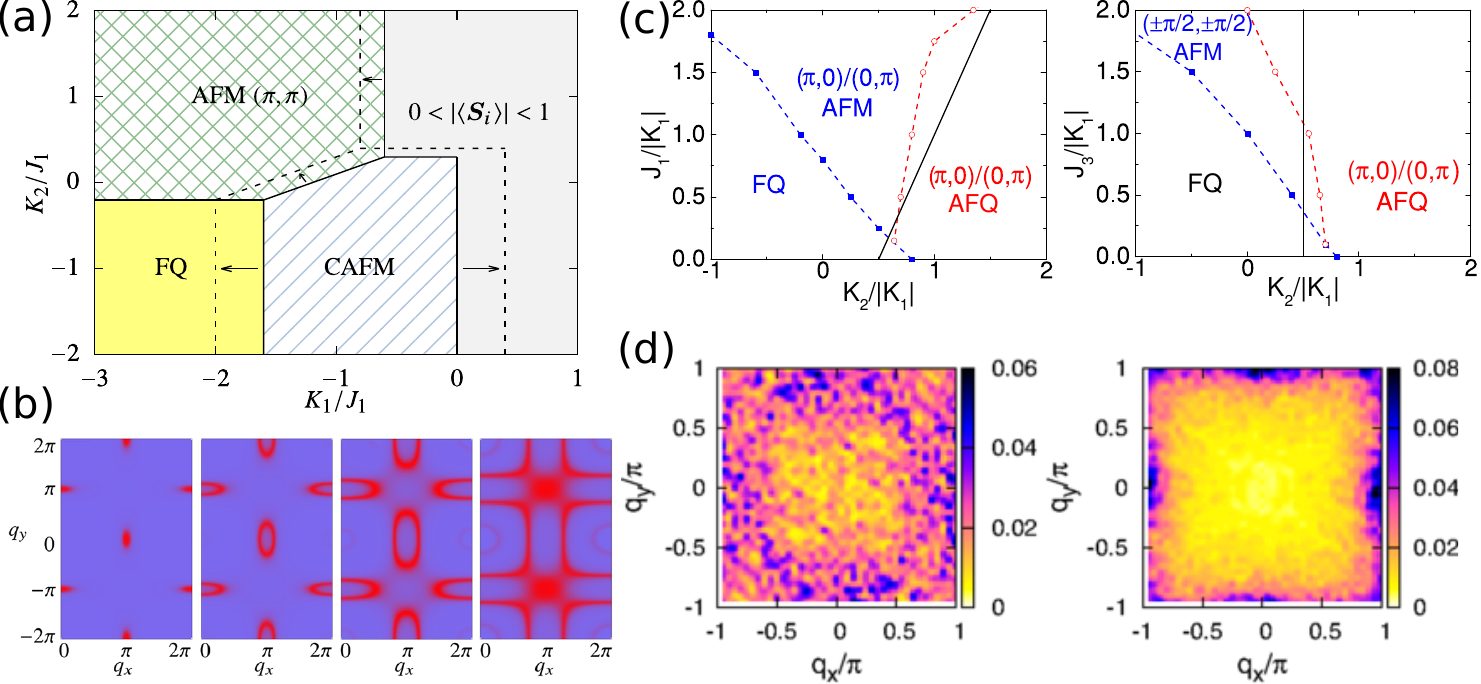}
	\caption{Ferroquadrupolar order and antiferroquadrupolar order as source for nematicity in FeSe. (a) Variational mean-field phase diagram of an Hamiltonian of localized $S=1$ spins with bilinear Heisenberg interactions $J_i$ and biquadratic interactions $K_i$ showing a ferroquadrupolar order (FQ), antiferromagnetic N\'eel order (AFM) and a columnar antiferromagnetic order (CAFM). The phase boundaries shift when breaking the $C_4$ symmetry of the model Hamiltonian by hand. (b) Expected dynamical structure factor in the FQ phase at energies $\omega/J_1=\{2,4,6,8\}$ (left to right). (c) Classical phase diagram of a bilinear-biquadratic Heisenberg model with antiferroquadrupolar order. The calculation is at finite temperature $T/|K_1|=0.01$ and the dashed lines show the crossovers between different orders (AFQ: antiferroquadrupolar order). (d) Expected momentum distribution of the dipolar magnetic structure factor in the presence of quadrupolar order for $K_2=-1$ (left) and $K_2=1.5$ (right) with $J_1=J_2=1$ and $K_1=-1$ as calculated within a classical Monte Carlo approach at $T/|K_1|=0.01$. (a), (b) Reproduced with permission from Ref. \cite{Wang_Z_16}, copyright 2016 American Physical Society. (c), (d)  Reproduced with permission from Ref. \cite{Yu2015}, copyright 2015 American Physical Society.}
	\label{fig:theory2}
\end{figure}

\begin{figure}[tb]
	\includegraphics[width=0.34\textwidth]{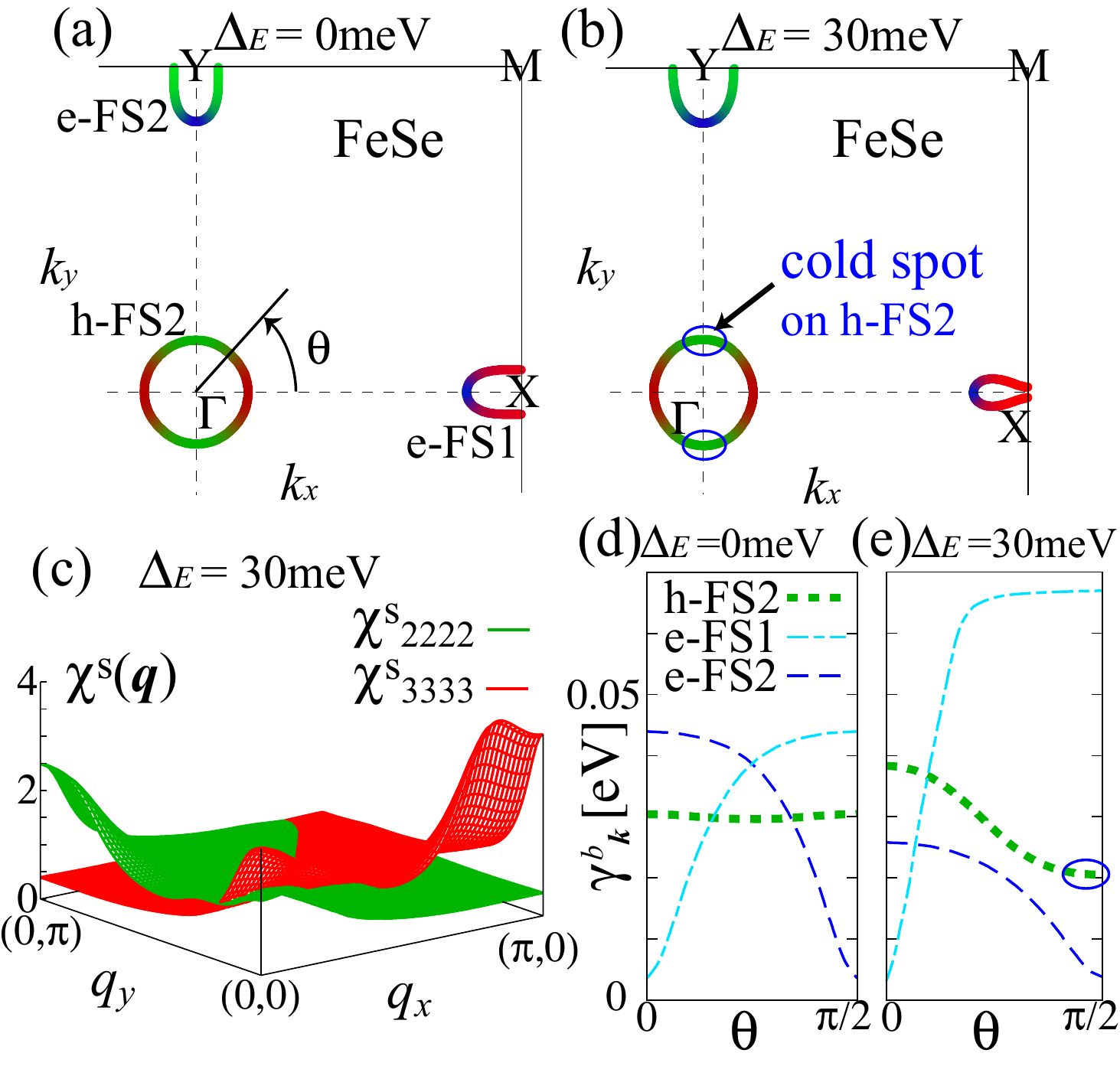}
	\includegraphics[width=0.65\textwidth]{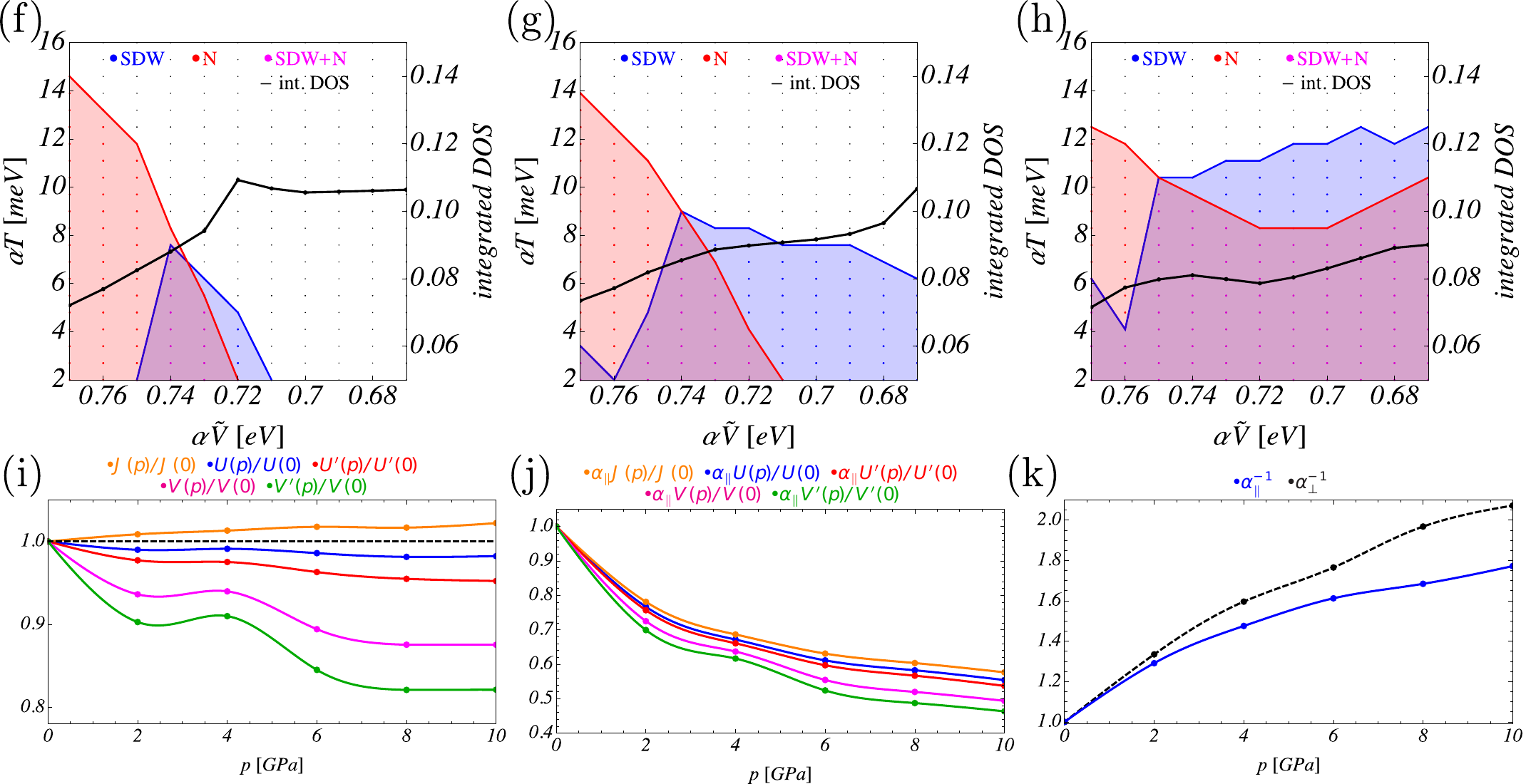}
	\caption{Transport anisotropy in FeSe \cite{Onari_2016arXiv} and orbital order from nearest neighbor Coulomb interactions \cite{Scherer_FeSe} (a) Fermi surface of a model for FeSe without orbital order and
	(b) Fermi surface of the same model with a shift of $\Delta E=30\,\mathrm{meV}$ where the Fermi surface at the X-point has the bowtie shape and the Fermi surface at the $\Gamma$ point exhibits cold spots
	at positions where few spin-fluctuations produce a damping of the quasiparticles. (c) Momentum structure of the spin-susceptibilities of the $d_{xz}$ and $d_{yz}$ orbitals with orbital order. (d) Quasiparticle
	damping $\gamma^b_k$ from a self-consistent calculation of the self-energy as a function of angle [see (a)] around the Fermi surfaces without orbital order and (e) the same with orbital order exhibiting a
	pronounced minimum at $\pi/2$ for the holelike Fermi surface, thus a cold spot [see (b)].
	(f-h) Phase diagram for FeSe as calculated from a mean field approach including nearest neighbor Coulomb interactions\cite{Scherer_FeSe}. Shown are the nematic ``N'' and magnetic ``SDW'' phases as function of temperature $T$ and interaction
	strength $\alpha \tilde V$ which decreases as function of pressure. The magnetic instability becomes stronger on increasing the rescaled Hund's coupling $\alpha J$: (f) $\alpha J=0.325$, (g) $\alpha J=0.35$, (h) $\alpha J=0.375$.
	The increase of the integrated density of states is consistent with the increase of the critical temperature for the superconducting instability as function of pressure as seen experimentally\cite{Medvedev2009,Mizuguchi2008,Margadonna2009}.
	(i) Pressure-induced renormalization of on-site interactions ($U$,$U'$,$J$) and extended Coulomb interactions ($V$, $V'$) extracted from DFT calculations relative to the values at ambient pressure\cite{Scherer_FeSe}. The Hund's coupling $J$ increases with pressure, while
	all other interactions show a downward trend. (j) The rescaled couplings with respect to the renormalization of the hoppings $\alpha_{\parallel}^{-1}$ (k) allows us to connect the rescaled interactions $\alpha \tilde V$ with the pressure in the phase diagrams (f-h).
	(a)-(d) Reproduced with permission from Ref. \cite{Onari_2016arXiv}, copyright 2017 American Physical Society. (f)-(k) Reproduced with permission from Ref. \cite{Scherer_FeSe}, copyright 2017 American Physical Society. }
	\label{fig:theory3}
\end{figure}

\subsection{Nematicity and magnetic fluctuations in FeSe}
Upon lowering temperature, many solid state systems undergo transitions to phases
with lower symmetries. FeSe, specifically, undergoes a transition to a nematic state $T_\mathrm{s}\sim 90$ {K}. However, unlike other iron based superconductors, FeSe does not develop magnetic order at low temperatures, which led to a strong debate concerning the origin of nematic order in FeSe. 
Possible sources for the nematic order could be lattice vibrations, spin fluctuations
or orbital fluctuations\cite{Fernandes2014}. In the case of FeSe, there are no experimental evidences for a lattice driven instability, but the other two scenarios have been proposed
for various reasons. The absence of strong spin fluctuations around $T_{\mathrm{s}}$ led to the conclusion of orbitally driven nematicity\cite{Baek2015}. Recent investigations on the spin excitations
via inelastic neutron scattering\cite{Wang2015} put further constraints to theoretical
models, namely, the transfer of spectral weight from N\'eel to stripe fluctuations and the emergence of low-energy spin fluctuations upon lowering the temperature.

Despite the fact that FeSe is a metal, Heisenberg models of localized spins have been applied to understand its properties. Ab initio studies suggest that the electronic structure of FeSe leads to various competing magnetic states and ultimately does not allow for long range magnetic order. As a result, pure nematic order can be realized down to low temperatures\cite{Glasbrenner2015,Liu_16,Busemeyer_16}. This scenario of competing magnetic state is also consistent with the pinned stripe-type charge ordering found in thin films of FeSe\cite{Li2017}.

\begin{figure}[tb]
	\includegraphics[width=\textwidth]{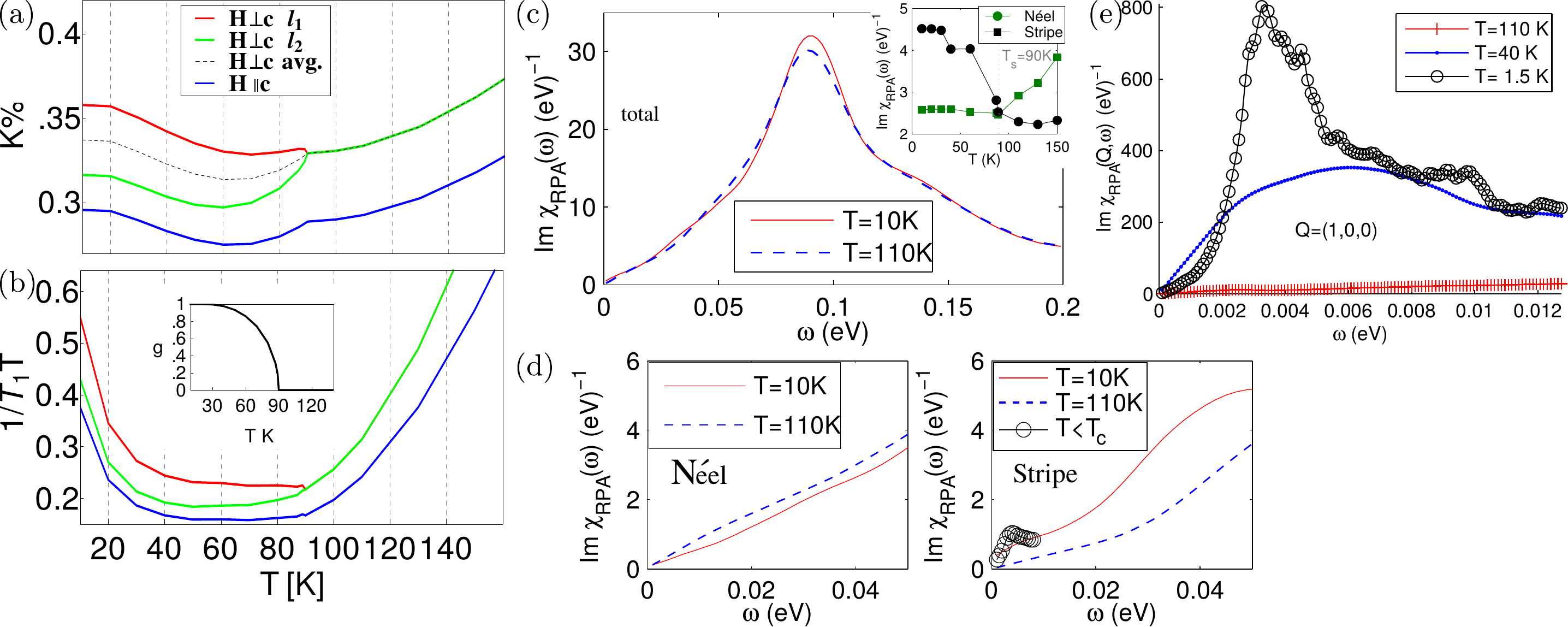}
	\caption{Spin fluctuations in FeSe from an itinerant model (a) Calculation of the Knight shift as expected in an orbital order scenario, (b) expected temperature dependence of the low energy spin fluctuations as measured in an NMR experiment via the spin-lattice relaxation rate divided by temperature $1/T_1T$. (c) The local susceptibility $\chi(\omega)$ is almost unchanged when entering the nematic phase, while the relative weight of the stripe and N\'eel fluctuations is reversed (inset) as seen experimentally\cite{Wang2015,WangLee2015NatPhys_FeSe-Para-Nematicity}, see Fig. \ref{fig:8}. (d) Integrated spin fluctuations of N\'eel type and stripe type at low energies from an itinerant model with orbital order. (e) Calculated spin resonance from sign-changing superconducting order parameter. (a), (b) Reproduced with permission from Ref. \cite{Mukherjee2015}, copyright 2015 American Physical Society. (c)-(e) Reproduced with permission from Ref. \cite{Kreisel15}, copyright 2015 American Physical Society.}
	\label{fig:theory3a}
\end{figure}

\begin{figure}[tb]
	\includegraphics[width=0.31\textwidth]{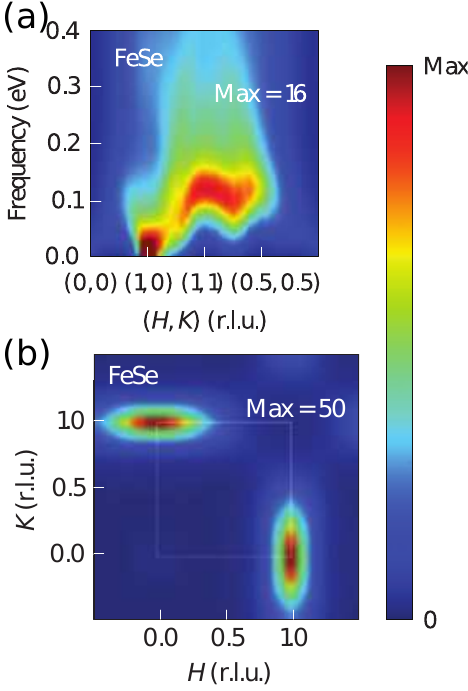}
	\includegraphics[width=0.68\textwidth]{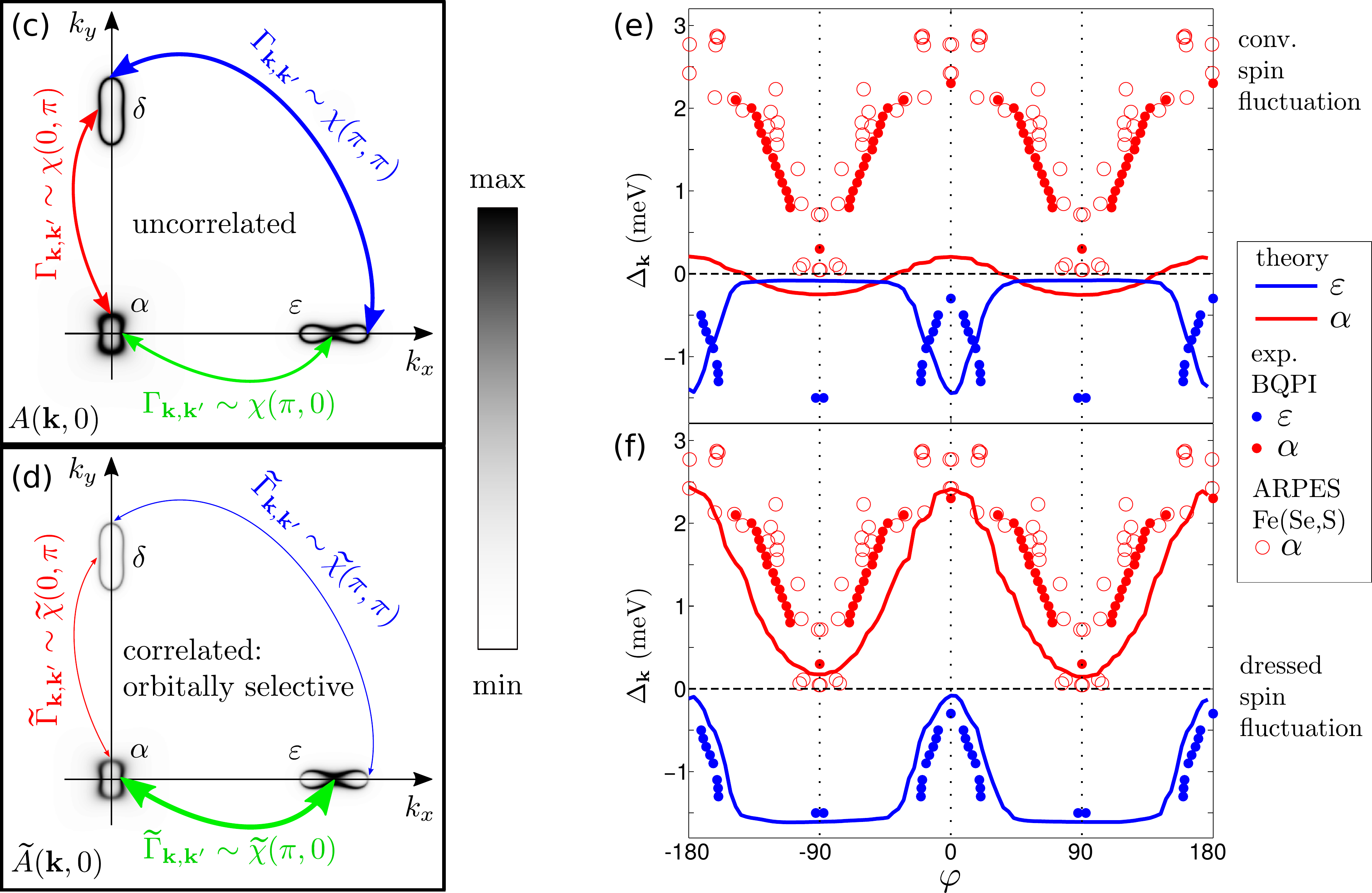}
	\caption{Spin fluctuations from a DMFT approach and superconducting pairing with orbital selective mechanism
	(a) Dynamic spin structure factor $S(q,\omega)$ calculated using DFT+DMFT, plotted along high symmetry directions. (b) Dynamic structure factor in the 2D plane $(H,K)$ at $k_z=\pi$. (c) Spectral function $A(k,\omega)$ of a fully coherent electronic structure for FeSe. A superconducting pairing from spin-fluctuations in this uncorrelated model leads to 3 almost equally dominant pairing interactions in the $d_{xy}$ orbital (blue), the $d_{xz}$ orbital (red) and the $d_{yz}$ orbital (blue). (d) Imposing orbital selectivity within modified quasiparticle weights changes the spectral function $\tilde A(k,\omega)$such that the $\delta$ pocket at $(0,\pi)$ becomes almost invisible and strongly suppresses the pairing in two orbitals leaving the dominating pairing glue $\tilde \Gamma_{k,k'}$ from the $d_{yz}$ orbital (thick, green arrow). (e) A conventional spin-fluctuation pairing mechanism with the model from (c) leads to a superconducting gap incompatible with experimental observations. (f) Including correlations via reduced quasiparticle weights and a splitting of those due to nematicity makes the pairing in the $d_{yz}$ orbital dominant\cite{Kreisel2017} and can account for the experimentally observed anisotropy\cite{Sprau2017}.
(a), (b) Reprinted by permission from Macmillan Publishers Ltd: Nature Physics, Ref. \cite{Yin2014}, copyright 2014. (c)-(f) Reproduced with permission from Ref. \cite{Kreisel2017}, copyright 2017 American Physical Society.}
	\label{fig:theory4}
\end{figure}

In other approaches to calculate the magnetic spectrum, the effects of band renormalizations are taken into account by starting from electronic structure models consistent with experimentally measured eigenenergies\cite{Kreisel15,Mukherjee2015} (Fig. \ref{fig:theory3a}).
The magnetic fluctuations have also been calculated by ab initio methods in conjunction with DMFT\cite{Yin2014,Skornyakov2017}, see Fig. \ref{fig:theory4} (a-b).
Including further neighbor interactions for localized spins, it has been argued that the nematic state is realized from quantum paramagnetic phase\cite{WangLee2015NatPhys_FeSe-Para-Nematicity}.
When biquadratic interactions are considered in a scenario driven by localized spins, the nematic properties of FeSe and spin-fluctuations seen in inelastic neutron scattering experiments were explained by a quadrupolar phase\cite{WangLee2015NatPhys_FeSe-Para-Nematicity,Yu2015,Wang_Z_16,Gong_2016arXiv160600937G,Hu_2016arXiv160601235H} and have signatures of dipolar fluctuations, see Fig. \ref{fig:theory2}.

The phase diagram of FeSe has been analyzed theoretically with mean-field approaches \cite{Glasbrenner2015}, field-theoretical methods\cite{WangLee2015NatPhys_FeSe-Para-Nematicity} and numerical  methods \cite{Yu2015,Wu_2016arXiv160302055W,Gong_2016arXiv160600937G,Wang_Z_16,Hu_2016arXiv160601235H}. 
A complementary microscopic approach starts from an itinerant model including a multiorbital electronic structure together with a Hubbard-Hund interaction. A variety of possible sources for nematic order in FeSe are revealed.
{\red A renormalization group analysis reveals that interaction parameters of different orbital components scale differently. Thus, magnetic fluctuations that lead to a superconducting instability also promote nematic order without leading to a magnetically ordered state\cite{Chubukov2015,Chubukov2016,Xing17,Honerkamp_17,Classen17}, see Fig. \ref{fig:theory0}.}
Spin-fluctuations together with orbital mismatch of the electron and hole Fermi surfaces may also be at the origin of nematicity \cite{Fanfarillo_2017arXiv}. Experimental evidence obtained by electronic Raman scattering suggests a Pomeranchuk instability \cite{Massat2016, Gallais2016}. Finally,  orbital fluctuations due to vertex corrections, as illustrated in Fig. \ref{fig:theory1}, have been proposed \cite{Yamakawa_PRX_16,Onari2015}. Furthermore, the transport anisotropy in the nematic phase has been argued to be due to the positions of cold spots on the Fermi surface which are determined by the spin fluctuations, see Fig. \ref{fig:theory3}(a-e)\cite{Onari_2016arXiv}. 

In the orbital-order scenarios for nematicity, various types of order can be ruled out because their symmetries are not compatible with common experimental investigations\cite{Yi_15} such that a bond order parameter seems essential to describe observed band splittings.
The experimentally observed sign change of the orbital polarization\cite{Suzuki2015} has been argued to be due
to the positive feedback between the nematic orbital order and the spin susceptibility and calculated in a self-consistent approach\cite{Onari2016}.
Interatomic Coulomb interactions were found to renormalize the band structure and also lead to such an order parameter\cite{Jiang_16,Wu_2016arXiv160302055W}.
In addition, the increase of the superconducting transition temperature together with a reduction
of the tendency towards nematicity as function of pressure can be understood in the same scenario, by using a combination of model-based calculations with input from ab-initio approaches\cite{Scherer_FeSe}, see Fig. \ref{fig:theory3}(f-k).

\subsection{Superconducting pairing in FeSe}
Similar to the microscopic models for nematicity, two general starting points seem reasonable for
the description of the superconducting pairing glue in iron-based superconductors. Localized spin models generally lead to a sign changing order parameter via spin-fluctuations, whereas itinerant models can take advantage of fluctuations of various kinds to support the superconducting instability.
Orbital fluctuations supposedly lead to an order parameter without sign change that may account for the slow suppression of the critical temperature upon isoelectronic substitution\cite{UrataKontaniTanigaki2016PRB_FeSe-Co-doping-pairing}.
Pairing arising from a nematic quantum spin liquid has been used to calculate the superconducting order parameter recently\cite{She2017arXiv}.

For a superconducting instability in presence of nematicity  {\red (i.e. in a orthorhombic system)}, the common symmetry classification in s-wave and d-wave instabilities is not valid anymore; {\red this is a consequence which} is independent of the underlying pairing interaction. The resulting order parameter is of lower symmetry as well and can lead to accidental nodes in presence of a sign-changing order parameter\cite{Kreisel15,Mukherjee2015}. The role of the nematic order parameter for the pairing interaction is still under debate. Under the assumption that both instabilities compete, the critical temperature should increase when nematicity is suppressed\cite{Baek2015,Watson2015III}. 
It turns out that nematic fluctuations are capable of pairing electrons\cite{Kang_2016arXiv160601170K}. Finally, it is possible that more than one single pairing mechanism is responsible for the various properties as it has been suggested by the observation of a double dome in doped FeSe films\cite{Song_16}.

Electronic correlations play an important role in iron-based superconductors in general and are supposed to be large in FeSe in particular\cite{Maletz2014,Watson2015,Watson2016,Yin2011}. As already discussed in many respects for iron-based superconductors, electronic states dominated by some orbitals are more strongly correlated than states dominated by other orbitals. Describing the electronic structure within a Fermi-liquid picture, this leads to substantial differences in quasiparticle weights and interactions. This orbital selectivity influences properties of the magnetism and allows for orbital ordering\cite{deMedici_review,Bascones_review,Biermann_review,Yin2011,Li_orb_sel_16,Haule16,Ye_doping_FeSc_14}. Indeed, the Cooper pairing itself can also become orbital-selective, thus electrons of a specific orbital character form Cooper pairs of the superconductor yielding a highly anisotropic superconducting energy gap \cite{Ogata_selectivepairing,Si_selectivepairing}. The consequences of these orbital selective electronic correlations for superconductivity in FeSe were discussed recently\cite{Sprau2017,Kreisel2017}. Implementing this idea in a phenomenological approach describes the experimentally observed variations of the superconducting energy gap in FeSe quite well when assuming a spin-fluctuation driven pairing interaction\cite{Sprau2017,Kreisel2017} [Fig. \ref{fig:theory4}(c-f)]. At the same time, also properties of the scattering on native impurities and trends in the magnetic fluctuation spectrum can be understood.


\section{Concluding remarks}

FeSe is host to a complex phase interplay between nematicity, magnetism and superconductivity, that at first sight appears very different from ``typical'' iron-based superconductors. 

The orthorhombic phase of FeSe that occurs over a wide temperature range, resembles the nematic phase that is found only in a narrow temperature range in many iron-pnictides. In both cases, this phase is reached via a second-order phase transition from the tetragonal high-temperature phase. The signatures at $T_\mathrm{s}$ in magnetic susceptibility, resistivity and specific heat in FeSe are reminiscent of the respective anomalies at $T_\mathrm{s}$ in those electron-doped BaFe$_2$As$_2$ systems that show a split between $T_\mathrm{s}$ and $T_\mathrm{N}$. Similarly, the nematic susceptibility, as inferred by various probes, behaves in a similar manner in FeSe and typical 122-type iron-arsenides. The elastic shear modulus and the electronic Raman responses are very similar as well. 
However, momentum-resolved probes show some possible differences in the nematic phase between the two types of systems. The magnetic fluctuation spectrum in FeSe is notably more complex than in many 122-type superconductors, with both stripe-type and checkerboard-type low-energy magnetic fluctuations. In addition, the change of the electronic band structure in crossing $T_\mathrm{s}$ might be more complex in FeSe than in the 122's, necessitating 'bond-type' nematic order for its explanation. Though they are often assumed to be equivalent, the relation between the nematic phase in FeSe and in 122-systems appears to be a central open question. Its resolution is complicated by the small temperature extent of the nematic phase in systems other than FeSe. 

Magnetic order in FeSe has been found only under applied pressure. The coupling of orthorhombic lattice distortion and magnetic order is, however, similar in FeSe and doped 122-type materials. When magnetic order sets in with $T_\mathrm{N}<T_\mathrm{s}$, the orthorhombic distortion increases abruptly in the magnetic phase, not dissimilar to the behavior of lightly Co- and Rh-doped BaFe$_2$As$_2$. Moreover, the enhancement of stripe-type magnetic fluctuations in the orthorhombic phase is reminiscent of the familiar iron-arsenide systems. Finally, magnetic order and orthorhombic distortion set in as a joint first-order phase transition in pressurized FeSe, similar to the case in many (hole-doped or isovalently substituted) 122-type iron-based superconductors. 

Whereas structure and magnetism in substituted BaFe$_2$As$_2$ and pressurized FeSe appear analogous, the coupling between superconductivity and nematic/magnetic order remains enigmatic in FeSe. The onset of superconductivity has only a minor effect on nematicity in FeSe, whereas in sulfur-substituted FeSe, superconductivity even appears to favor a larger orthorhombic distortion - in contrast to all other known iron-based systems. Finally, it remains an open question whether superconductivity coexists with the pressure-induced magnetic order.

The realization of a nematic state in FeSe allows to test and possibly falsify theoretical scenarios and mechanisms for nematicity and its interplay with other phases in this system. Itinerant models of the electronic structure and models of localized magnetic moments have been starting points for theoretical investigations. Ab-initio based methods are able to calculate important quantities, but model-based calculations are still needed to understand the basic principles and put FeSe and its properties in the context of other iron-based materials. These approaches need input from experimental investigations and/or first-principle investigations to justify the theoretical starting point. For the superconducting state it seems that new mechanisms are required to explain the exotic properties of FeSe at ambient pressure and understand the phase interplay in pressurized FeSe.

The study of FeSe has revealed a fascinating variant of the interplay between structure, magnetism and superconductivity in iron-based systems. In particular, the orthorhombic distortion, or nematicity, has been firmly established as an independent player that may be distinct from magnetism. Future surprising discoveries, possibly in doped or intercalated FeSe-based systems seem likely. The complexity of the FeSe phase diagram when chemical and physical pressure are combined may be a hint at yet undiscovered phenomena.

Acknowledgments: We would like to thank Brian M. Andersen, Andreas Kreyssig, William R. Meier and Aashish Sapkota for interesting and useful discussions. The work at Ames Laboratory was supported by the US Department of Energy under DE-AC02-07CH11358.


\end{document}